\documentclass{iucrjournals}

\usepackage{hyperref}
\hypersetup{
    colorlinks=true,
    linkcolor=blue,
    filecolor=magenta,      
    urlcolor=cyan,
    citecolor=magenta
}

\usepackage{dsfont} 
\usepackage{amsfonts}
\usepackage{amsthm}
\usepackage{amsmath} 

\newtheorem{theorem}{Theorem}[section]
\newtheorem{corollary}{Corollary}[theorem]

\usepackage{algorithm}
\usepackage{algpseudocode}

\title{PETIMOT: A Novel Framework for Inferring Protein Motions from Sparse Data Using SE(3)-Equivariant Graph Neural Networks}

\author[a]{Valentin Lombard
}
\author[a]{Julien Nguyen Van
}
\author[c]{
Sergei Grudinin 
\IUCrCemaillink{Sergei.Grudinin@univ-grenoble-alpes.fr}
\IUCrOrcidlink{0000-0002-1903-7220}
\IUCrAufn{These authors jointly supervised the work.}
}
\author[a,b]{
Elodie Laine
\IUCrCemaillink{elodie.laine@sorbonne-universite.fr}
\IUCrOrcidlink{0000-0003-4870-6304}
\IUCrAufn[1]{} 
}

\affil[a]{Department of Computational, Quantitative, 
and Synthetic Biology (CQSB), UMR 7238 
IBPS, Sorbonne Université, CNRS 
Paris, 75005, France }
\affil[b]{Institut universitaire de France (IUF) }
\affil[c]{Univ. Grenoble Alpes, CNRS, Grenoble INP, LJK 38000 Grenoble, France.  }

\begin{document} 
\maketitle 

\begin{synopsis}
We present a new formulation of protein flexibility and learn protein motions from sparse experimental data
\end{synopsis}

\begin{abstract}
Proteins move and deform to ensure their biological functions. Despite significant progress in protein structure prediction, approximating conformational ensembles at physiological conditions remains a fundamental open problem. This paper presents a novel perspective on the problem by directly targeting continuous compact representations of protein motions inferred from sparse experimental observations. We develop a task-specific loss function enforcing data symmetries, including scaling and permutation operations. Our method PETIMOT (Protein sEquence and sTructure-based Inference of MOTions) leverages transfer learning from pre-trained protein language models through an SE(3)-equivariant graph neural network. When trained and evaluated on the Protein Data Bank, PETIMOT shows superior performance  in time and accuracy, capturing protein dynamics, particularly large/slow conformational changes, compared to state-of-the-art diffusion and flow-matching approaches, as well as traditional physics-based models.
Our code and protocols are available at \url{https://github.com/PhyloSofS-Team/PETIMOT}.
\end{abstract}


\section{Introduction}
\label{intro}
Proteins orchestrate biological processes in living organisms by interacting with their environment and adapting their three-dimensional (3D) structures to engage with cellular partners, including other proteins, nucleic acids, small-molecule ligands, and co-factors. In recent years, spectacular advances in high-throughput deep learning (DL) technologies have provided access to reliable predictions of protein 3D structures at the scale of entire proteomes \cite{varadi2024alphafold}. These breakthroughs have also highlighted the complexities of protein conformational heterogeneity. State-of-the-art predictors struggle to model alternative conformations, fold switches, large-amplitude conformational changes, and solution ensembles \cite{chakravarty2025proteins}. 

The success of AlphaFold2 \cite{Jumper:2021rb} has stimulated machine-learning approaches focused on inference-time interventions in the model to generate structural diversity. They include enabling or increasing dropout \cite{brysbaert2024massivefold,wallner2023afsample}, or manipulating the evolutionary information given as input to the model \cite{kalakoti2024afsample2,wayment2023predicting,del2022sampling,stein2022speach_af}. Despite promising results on specific families, several studies have emphasised the difficulties in rationalising the effectiveness of these modifications and interpreting them \cite{porter2024metamorphic,bryant2024structure}. Moreover, these cannot be transferred to protein language model (pLM)-based predictors that do not rely on multiple sequence alignments. Researchers have also actively engaged in the development of deep learning frameworks based on diffusion, or the more general flow matching, to generate conformational ensembles \cite{lewis2025scalable,wang2403protein}. While being several orders of magnitude cheaper than Molecular Dynamics (MD) simulations, these models remain computationally intensive, require massive MD training data, and are limited to sampling approximate equilibrium distributions.

{This work proposes a new perspective on} the protein conformational diversity problem. Instead of learning and sampling from multi-dimensional empirical distributions, we propose to learn eigenspaces (the structure) of the positional covariance matrices in collections of experimental 3D structures and generalize these over different homology levels. The use of experimental structure collections to infer protein dynamics through Principal Component Analysis (PCA) is well-established in the literature \cite{best2006relation,schneider2025ensembleflex,lombard2024explaining,yang2009principal}. The diversity present within -- even a modest number of -- experimental 3D structures of the same protein or close homologs is a good proxy for the conformational heterogeneity of proteins in solution \cite{best2006relation} and can generally be (almost fully) explained by a small set of linear vectors, also referred to as modes \cite{lombard2024explaining,yang2009principal}. Moreover, interpolation trajectories performed in PCA space inferred from experimental structures can recapitulate intermediate functional states \cite{lombard2024explaining}. Although linear spaces may not be well-suited for capturing highly complex non-linear motions, such as loop deformations, they offer multiple advantages. These include faster learning due to the reduced complexity of the model, improved explainability as the components directly correspond to interpretable data dimensions, faster inference, and the straightforward combination or integration of multiple data dimensions.

To summarize, our main contributions are:

\begin{itemize}
    \item We provide a novel formulation of the protein conformational diversity problem.
    \item We present a novel benchmark representative of the Protein Data Bank structural diversity and compiled with a robust pipeline \cite{lombard2024explaining}, along with data- and task-specific metrics.
    \item We develop a SE(3)-equivariant Graph Neural Network architecture equipped with a novel symmetry-aware loss function for comparing linear subspaces, with invariance to permutation and scaling. Our model, PETIMOT, leverages embeddings from pre-trained pLMs, building on prior proof-of-concept work demonstrating that they encode information about functional protein motions \cite{lombard2024seamoon}.
    \item PETIMOT is trained on sparse experimental data without any use of simulation data, in contrast with Timewarp for instance \cite{klein2024timewarp}. Moreover, our model does not require physics-based guidance or feedback,  unlike \cite{wang2403protein} for instance.
    \item Our results demonstrate the capability of PETIMOT to generalise across protein families (contrary to variational autoencoder-based approaches) and to compare favorably in running time and accuracy to the physics-based Normal Mode Analysis.
\end{itemize}

\section{Related Works}

\paragraph{Protein structure prediction and generating conformational ensembles.} 
AlphaFold2 was the first end-to-end deep neural network to achieve near-experimental accuracy in predicting protein 3D structures, even for challenging cases with low sequence similarity to proteins with resolved structures \cite{Jumper:2021rb}. 
Later works have shown that substituting the input alignment by embeddings from a pLM can yield comparable performance \cite{lin2023evolutionary,hayes2024simulating,weissenow2022protein,wu2022high}.

Beyond the single-structure frontier, several studies have underscored the limitations and potential of protein structure predictors (PSP) for generating alternative conformations \cite{saldano2022impact,lane2023protein,bryant2024structure,chakravarty2025proteins}. Approaches focused on re-purposing AlphaFold2 include dropout-based massive sampling \cite{brysbaert2024massivefold,wallner2023afsample}, guiding the predictions with state-annotated templates \cite{faezov2023alphafold2,heo2022multi}, and inputting shallow, masked, corrupted, subsampled or clustered alignments \cite{kalakoti2024afsample2,wayment2023predicting,del2022sampling,stein2022speach_af}. 
Despite promising results, these approaches remain computationally expensive and their generalisability, interpretability, and controllability remain unclear \cite{bryant2024structure,chakravarty2025proteins}. More recent works have aimed at overcoming these limitations by directly optimising PSP learnt embeddings under low-dimensional ensemble constraints \cite{yu2025esmadam}

Another line of research has consisted in fine-tuning or re-training AlphaFold2 and other single-state PSP under diffusion or flow matching frameworks \cite{jing2024alphafold,abramson2024accurate,krishna2024generalized}. 
More generally, diffusion- and flow matching-based models allow for efficiently generating diverse conformations conditioned on the presence of ligands or cellular partners \cite{jing2023eigenfold,ingraham2023illuminating,wang2403protein,liu2024design}. Despite their strengths, these techniques are prone to hallucination.

{The recent availability of large-scale molecular dynamics (MD) datasets \cite{vander2024atlas, siebenmorgen2024misato,mokhtari2026dynarepo} has opened the door to a parallel line of research training deep learning models directly on these data to enhance or replace MD exploration. A first group of methods develops machine-learning force fields based on equivariant GNN representations \cite{wang2024ab}. A second directly emulates MD trajectories as generative tasks, enabling forward simulation, transition path sampling, and trajectory upsampling \cite{jing2024generative,costa2024equijump}. A third learns generative models of equilibrium Boltzmann distributions \cite{noe2019boltzmann,klein2024timewarp,zheng2024predicting,lewis2025scalable}. For instance, the BioEmu model \cite{lewis2025scalable}, trained on more than 200 ms of MD simulations and fine-tuned on experimental protein stability measurements, approximates equilibrium conformational distributions at a fraction of the cost of MD simulations, while capturing biologically meaningful conformational changes deposited in the PDB \cite{berman2000protein}.}

\paragraph{Protein conformational heterogeneity manifold learning.} 
Unsupervised, physics-based Normal Mode Analysis (NMA) has long been effective for inferring functional modes of deformation by leveraging the topology of a single protein 3D structure \cite{grudinin2020predicting,hoffmann2017nolb,hayward1995collective}. While appealing for its computational efficiency, the accuracy of NMA strongly depends on the initial topology \cite{laine2021hopma}, limiting its ability to model extensive secondary structure rearrangements. Recent efforts have sought to address these limitations by directly learning continuous, compact representations of protein motions from sparse experimental 3D structures. These approaches employ dimensionality reduction techniques, from classical manifold learning methods \cite{lombard2024explaining} to neural network architectures like variational auto-encoders \cite{ramaswamy2021deep}. By projecting motions onto a learned low-dimensional manifold, these methods enable reconstruction of accurate, physico-chemically realistic conformations, both within the interpolation regime and near the convex hull of the training data \cite{lombard2024explaining}. Additionally, they assist in identifying collective variables from molecular dynamics (MD) simulations, supporting importance-sampling strategies \cite{chen2023discovering,belkacemi2021chasing,bonati2021deep,wang2020machine,ribeiro2018reweighted}. Despite these advances, such approaches are currently constrained to family-specific models.

\paragraph{E(3)-equivariant graph neural networks.}
Graph Neural Networks (GNN) have been extensively used to represent protein 3D structures. They are robust to transformations of the Euclidean group, namely rotations, reflections, and translations, as well as to permutations. In their simplest formulation, each node represents an atom and any pair of atoms are connected by an edge if their distance is smaller than a cutoff or among the smallest $k$ interatomic distances. Many works have proposed to enrich this graph representation with   SE(3)-equivariant features informing the model about interatomic directions and orientations \cite{ingraham2019generative, jing2020learning, dauparas2022robust,krapp2023pesto,wang2024enhancing}. 
To go beyond local 3D neighbourhoods while maintaining sub-quadratic complexity, Chroma adds in randomly sampled long-range connections \cite{ingraham2023illuminating}. 





\section{Data representation and problem formulation}
\label{Data}

To generate training data, we exploit experimental protein single chain structures available in the PDB. We first clustered these chains based on their sequence similarity. Then, within each cluster, we aligned the protein sequences and used the resulting mapping for superimposing the 3D coordinates \cite{lombard2024explaining}. It may happen that some residues in the multiple sequence alignment do not have resolved 3D coordinates in all conformations. To account for this uncertainty, we assigned a confidence score $w_i$ to each residue $i$ computed as the proportion of conformations including this residue. The 3D superimposition puts the conformations' centers of mass to zero and then aims at determining the optimal least-squares rotation minimizing the Root Mean Square Deviation (RMSD) between any conformation and a reference conformation, while accounting for the confidence scores \cite{Kabsch:a12999,kearsley1989orthogonal},
\begin{equation}
E = \frac{1}{\sum_i w_i} \sum_i w_i (\vec r_{ij} - \vec r_{i0})^2,
\end{equation}
where $\vec r_{ij} \in \mathbb{R}^3$ is the $i$th centred coordinate of the $j$th conformation and $\vec r_{i0} \in \mathbb{R}^3$ is the $i$th centred coordinate of the reference conformation. Next, we defined our ground-truth targets as eigenspaces of the coverage-weighted C$\alpha$-atom positional covariance matrix,
\begin{equation}
C = \frac{1}{m-1} W^{\frac{1}{2}} R^c (R^c)^T W^{\frac{1}{2}} = \frac{1}{m-1} W^{\frac{1}{2}} (R-R^0) (R-R^0)^T W^{\frac{1}{2}},
\label{eq:covariance}
\end{equation}
where $R$ is the $3N \times m$ positional matrix with $N$ the number of residues and $m$ is the number of conformations, $R^0$ contains the coordinates of the reference conformation,  and $W$ is the $3N \times 3N$ diagonal coverage matrix. The covariance matrix is a $3N\times 3N$ square matrix, symmetric and real.  
We decompose $C$ as $C=YDY^T$, where $Y$ is a $3N\times 3N$ matrix with each column defining a coverage-weighted eigenvector or a principal component that we interpret as a {\em linear motion}. $D$ is a diagonal matrix containing the eigenvalues. The latter highly depend on the sampling biases in the PDB and thus we do not aim at predicting them. 

\paragraph{Problem formulation.}
For a protein of length $N$, let 
$Y$ be $3N \times K$  {\em orthogonal} ground-truth deformations, 
\begin{equation}
    Y^T Y =  I_{K}.
\end{equation}

Our goal is to find coverage-weighted vectors 
$X \in \mathbb R^{3N \times L}$
whose components $l$ {\em approximate} some components $k$ of the ground truth $Y$:
\begin{equation}
    W^{1\over{2}} \mathbf{\tilde x}_l  \approx \mathbf y_k.
\end{equation}

Below, we provide three alternative formulations for this problem.
%
%
PETIMOT's loss function serves two key purposes: it enables effective training of the network to predict subspaces representing  multiple distinct modes of deformations -- {\it i.e.}, with low overlap between the subspace's individual linear vectors, while preventing convergence to a single dominant mode. 

\paragraph{The least-square formulation.}
To evaluate a predicted motion direction against a ground-truth direction, we use a Least-Square (LS) error, which, together with Mean Absolute Error (MAE), is among the most accepted metrics for regression tasks. Here, we have specifically adapted it to the challenge of evaluating directional motion vectors rather than static coordinates, and scaled between 0 and 1 for better training, interpretability and usability.

For each protein of length $N$ with a coverage $W$, we compute the weighted pairwise {\em least-square difference} $\mathcal{L}_{kl}$ between ground-truth directions $Y$ and predicted motion directions $ X$ for each pair of a $k$ direction in the ground truth and an $l$ direction in the prediction as,
\begin{equation}
\begin{split}
\mathcal{L}_{kl} =& \frac{1}{N} \sum_{i=1}^N \|\vec{y}_{ik} -  w_i^{1/2} c_{kl}\vec{ x}_{il}\|^2
=
\frac{1}{N} \mathbf{y}_k^T  \mathbf{ y}_k
- \frac{1}{N}  
\frac{
(\mathbf{y}_{k}^T W^{1\over 2} \mathbf{ x}_{l})^2
}{
\mathbf{ x}_{l}^T W \mathbf{ x}_{l}
}
,
\end{split}
\end{equation}
where we scaled the ground-truth tensors such that $Y^TY = N I_K$ and we used the fact that the optimal scaling coefficients   $c_{kl}$   between the $k$-th ground truth 
vector
and the $l$-th prediction are given by
\begin{equation}
c_{kl} = \frac{
\sum_{i=1}^N w_i^{1\over 2} \mathbf{y}_{ik}^T \mathbf{ x}_{il} }
{ \sum_{i=1}^N w_i \mathbf{ x}_{il}^T\mathbf{ x}_{il}}
=
\frac{
\mathbf{y}_{k}^TW^{1\over 2} \mathbf{ x}_{l}
}{
\mathbf{ x}_{l}^T W \mathbf{ x}_{l}
}
.
\end{equation}

This invariance to global scaling is motivated by the fact that we aim at capturing the relative magnitudes and directions of the motion patterns rather than their sign or absolute amplitudes. 

\paragraph{Linear assignment problem.} 
We then formulate an {\em optimal linear  assignment problem} to find the minimum-cost matching between the ground-truth and the predicted directions. Specifically, we aim to solve the following assignment problem for the least-square (LS) costs,
\begin{equation}
    \begin{split}
&\text{LS Loss} = \frac{1}{\min(K, L)}
\min_{\pi \in S_{J}} \sum_{k=1}^{\min(K, L)} \mathcal{L}_{k,\pi(k)}\\
%
\text{subject to:}~~~~~
&\pi : \{1,\ldots,\min(K, L)\} \rightarrow \{1,\ldots,L\},~~
\pi(k) \neq \pi(k') \text{ for } k \neq k',
\end{split}
\end{equation}
where $K$ and $L$ are the number of ground-truth and predicted directions respectively, 
and $\pi(k)$ represents the index of the predicted direction assigned to the $k$-th ground truth direction.
This formulation ensures an optimal one-to-one matching, while accommodating cases where the number of predicted and ground-truth directions differs. We backpropagate the loss only through the optimally matched pairs, using {\em scipy} linear\_sum\_assignment. We have also tested a smooth version of the loss above with continuous gradients, but it did not improve the performance.

\paragraph{The subspace coverage formulation.} 
We propose another formulation of the problem in terms of the subspace coverage metrics \cite{amadei1999convergence,leo2005analysis,david2011characterizing}.
Specifically, we sum up {\em squared sinus} (SS) dissimilarities between ground-truth and predicted directions (formally computed as one minus squared cosine similarity),
\begin{equation}
 \text{SS Loss} = 1 - \frac{1}{\min(K,L)} \sum_{k=1}^{K} \sum_{l=1}^{L} ( {\mathbf y}_{k}^T W^{1\over 2} {\mathbf x^{\perp}_{l}} )^2,
\end{equation} 
 where the subspace $\{\mathbf{x}^{\perp}_{l}\}$ is obtained by orthogonalising the coverage-weighted predicted linear subspace $\{ W^{1\over 2}\mathbf{x}_{l}\}$, where $\mathbf x_l ^T W \mathbf x_l = 1$, using the Gram–Schmidt process.
 This operation ensures that the loss ranges from zero for identical subspaces
 to one for 
 mutually orthogonal subspaces 
 and avoids artificially inflating the SS loss due to redundancy in the predicted motions. The order in which the predicted vectors are orthogonalised does not influence the loss, guaranteeing stable training.
Appendix \ref{app:loss:invariance} proves this statement.
The SS loss is conceptually similar to the comparison of angles between subspaces -- see a few recent examples of such subspace comparison from other ML domains in \cite{zhu2021class,feng2023fair,chen2023inner,hawke2024contrastive,schlaginhaufen2024towards}.

\paragraph{Independent Subspace (IS) Loss.}
We can substitute the orthogonalisation procedure by using an auxiliary loss component for maximising the rank of the predicted subspace.
For this purpose, we chose the squared cosine similarity computed between pairs of predicted vectors.
The final expression for the {\em independent subspace} (IS) loss is
\begin{equation}
 \text{IS Loss} =  
 \frac{1}{L^2} \sum_{l=1}^{L} \sum_{m=1}^{L}  ( \textbf{x}_{l}^T W \textbf{x}_{m} )^2
 - 
 \frac{1}{L\min(K,L)} \sum_{k=1}^{K} \sum_{l=1}^{L} ( \textbf{y}_{k}^T W^{\frac{1}{2}} \textbf{x}_{l} )^2
 ,
\end{equation}
where the predictions $\{\mathbf{x}_l\}$ are normalised prior to the loss computation such that $\mathbf x_l ^T W \mathbf x_l = 1$ and the scaling factors ensure that the loss ranges between 0 and 1. 
Appendix \ref{app:loss:invariance} analyses the stability of this formulation.

\section{Architecture}

\begin{figure}[h]
\centering
\includegraphics[width=1\linewidth]{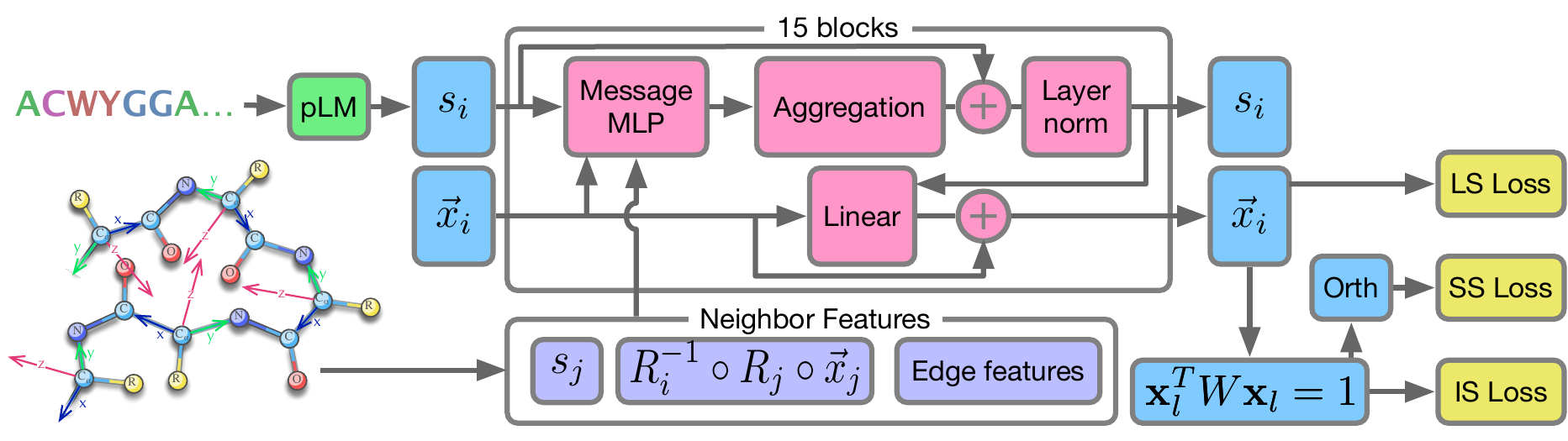}
\caption{{\bf PETIMOT's architecture overview.} The model processes sequence embeddings ($s$) and motion vectors ($\vec{x}$) through 15 message-passing blocks. The GNN topology and edge features are defined from the input 3D coordinates. Edge features encode 3D geometrical properties such as the relative spatial relationships between residue pairs. Each block updates both $s$ and $\vec{x}$ representations by aggregating information from neighboring residues.  SE(3) equivariance is achieved by computing the features of neighbors $j$ in the reference frame of the central residue $i$. 
Three types of losses (LS, SS, and IS) are computed,
with prior normalization of the predictions for the IS and SS losses, and an additional orthogonalisation of the predictions for the SS loss.}
\label{fig:architecture}
\end{figure}

We solve the problem formulated above with a pLM-informed SE(3)-equivariant graph neural network called PETIMOT (Fig. \ref{fig:architecture}). PETIMOT takes as input a protein sequence of length $N$, converted into an embedding $\mathbf{s}$ by a pre-trained pLM, along with 3D coordinates, and outputs a set of linear motions $X \in \mathbb R^{3N \times L}$.

\paragraph{Dual-track representation.} PETIMOT processes protein sequences through a message-passing neural network that simultaneously handles residue embeddings and motion vectors in local coordinate frames (Fig. \ref{fig:architecture}). For each residue $i$, we define and update a node embedding $\mathbf{s}_i \in \mathbb{R}^d$ initialized from pLM features and a set of $K$ motion vectors $\{\vec{x}_{ik}\}_{k=1}^K \in \mathbb{R}^{3\times K}$ initialized randomly. The message passing procedure is detailed in Algorithm \ref{alg:PETIMOT} of Appendix \ref{MPBlockA}. The protein is represented as a graph where nodes correspond to the residues, and edges capture spatial relationships. We connect each residue $i$ to its $k$ nearest neighbors based on C$\alpha$ distances in the input structure and $l$ randomly selected residues. This hybrid connectivity scheme ensures both local geometric consistency and global information flow, while maintaining sparsity for computational efficiency. 
Indeed, our model scales {\em linearly} with the length $N$ of a protein.
In our base model we set $k=5$ and $l=10$.
\paragraph{Node and edge features.} We chose ProstT5 as our default pLM for initialising node embeddings \cite{heinzinger2023prostt5}. This structure-aware pLM offers an excellent balance between model size -- including the number of parameters and embedding dimensionality -- and performance  \cite{lombard2024seamoon}. Each residue's backbone atoms (N, CA, C) define a local reference frame through a rigid transformation $T_i \in SE(3)$. For each residue pair $(i,j)$, we compute their relative transformation $T_{ij} = T_i^{-1} \circ T_j$ from which we extract the rotation $R_{ij} \in SO(3)$ and translation $\vec{t}_{ij} \in \mathbb{R}^3$. Under global rotations and translations of the protein, these relative transformations remain invariant. Edge features $e_{ij}$ provide an SE(3)-invariant encoding of the protein structure through relative orientations, translational offsets, protein chain distance, and a complete description of peptide plane positioning captured by pairwise backbone atom distances. See Appendix \ref{featuresA} for more details.

\section{Results}

\begin{figure}[ht!]
    \centering
    \includegraphics[width=\linewidth]{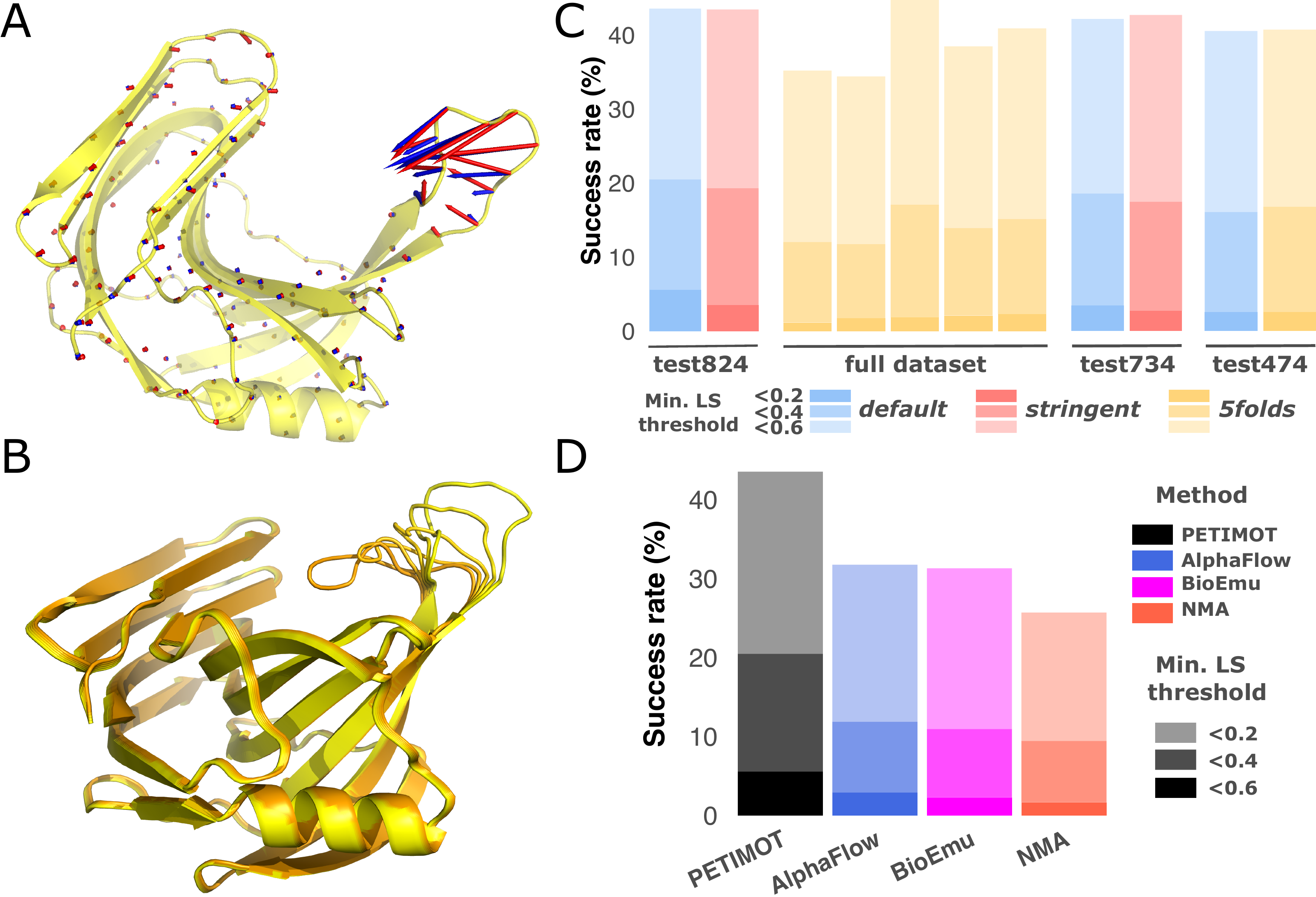}
    \caption{{{\bf PETIMOT prediction visualisation, evaluation and comparison with other methods.} {\bf A-B.} Prediction for {\it B. subtilis} xylanase A (PDB id: 3EXU, chain A). {\bf A.} The predicted and ground-truth vectors are in blue and red, respectively (LS$=0.15$). {\bf B.} Trajectory generated by deforming the protein structure along the predicted motion. {\bf C-D.} Success rates computed as the proportions of test proteins with Min. LS below 0.2 (dark), 0.4 (mild), and 0.6 (light). {\bf C.} Comparison of PETIMOT's \emph{default} (blue), \emph{stringent} (tomato red) and \emph{5folds} (gold) models. Test sets avoid data leakage at different levels. \emph{test824}: less than 80\% sequence similarity to any collection used for training. \emph{full dataset}: folds are defined based on strict sequence and structural similarity filters. \emph{test734}: less than 30\% sequence similarity to train collections. \emph{test474}: less than 30\% sequence similarity and no significant structural similarity to train collections. {\bf D.} Comparison of PETIMOT's \emph{default} model (black) with AlphaFlow (blue), BioEmu (magenta), and the NMA (tomato red) on \emph{test824}.}}
    \label{fig:indiv}
\end{figure}

\paragraph{Training and evaluation.}
We trained PETIMOT against linear motions extracted from all $\sim$750,000 protein chains from the PDB (as of June 2023) clustered at 80\% sequence identity and coverage. Our full training data comprises 7~335 conformational collections, which we augmented by computing the motions with respect to 5 reference conformations per collection. As a result, the full training set comprises 36,675 samples. This reference dataset encompasses conformations solved by multiple experimental techniques, including 56,866 Cryo-EM structures (30.5\%) and 2,187 NMR structures (about 1.5\%). This ensures representation of diverse conformational states beyond those accessible to X-ray crystallography. {Moreover, only 5.6\% of the training samples exhibit a maximum pairwise RMSD below 2 \AA{}, indicating that the vast majority of the dataset captures substantial conformational diversity.} We set the numbers of predicted and ground-truth motions, $K=L=4$. See Appendices \ref{trainingA} and \ref{ref:traning} for more details. At inference, we consider $w_i=1$, $\forall i=1..N$. We rely on four main evaluation metrics aimed at addressing the following questions:
{\bf 1)}    Is PETIMOT able to approximate at least one of the main linear motions of a given protein? For this, we rely on the minimum LS error over all possible pairs of predicted and ground-truth vectors. A prediction with LS$\leq0.2$ almost perfectly superimposes to the ground-truth motion {(Fig. \ref{fig:indiv}A-B)}. We consider predictions with LS$>0.6$ as inaccurate as they typically miss or indicate completely wrong directions for a large part of the residues involved in the motion. By comparison, the LS errors computed for random predictions are typically above 0.9.
{\bf 2)}    To what extent does PETIMOT capture the main motion linear subspace of a given protein? For this, we use the Global SS error;
{\bf 3)}    Is PETIMOT able to identify the residues that move the most? Here, we rely on the magnitude error, $ \frac{1}{N} \sum_{i=1}^N (\|\vec{y}_{ik}\|^2 -  \|c_{kl}\vec{ x}_{il}\|^2)$. 
{\bf 4)} Can PETIMOT be used to generate conformations resembling experimentally resolved functional protein states? For this, we generate conformations by deforming the input protein structure along the predicted motions and compute their RMSD to five diverse conformations selected from the ground-truth collection. See Appendix \ref{evaluation} for more details.

\paragraph{Robustness and generalisation.}

We tested PETIMOT's generalisation capabilities using three training protocols. In two of them, we randomly split the reference dataset into 70\% for training, 15\% for validation and 15\% for test, where any test protein has less than 80\% ({\it default}) or 30\% ({\it stringent}) sequence similarity to the train proteins. In addition, we conducted 5-fold cross-validation ensuring that each fold's training set did not contain any protein chain sharing significant structural or sequence similarity with the test set (\emph{5folds}). This protocol strictly prevents data leakage and provides robust evaluation across our complete dataset (Appendix \ref{ref:traning}).  PETIMOT's performance is robust and generalisable across the different data partitions, with success rates, defined as the fractions of test proteins with Min. LS$\leq0.6$, in the 35-45\% range (Fig. \ref{fig:indiv}C, Tables \ref{tab:additional} and \ref{tab:fullComp}). {Moreover, it showed limited sensitivity to the choice of reference conformation, predicting acceptable motions for at least 2 out of 5 reference conformations in 80\% of successful collections (Fig. \ref{supfig:influRef}a).}
See Appendix \ref{resultsA} for more details.

\paragraph{Biological relevance.} 
To assess the biological relevance of PETIMOT's predictions, we focused on three case studies: open-closed transitions, fold switches, and multi-state cryo-EM resolved structures. For open-closed transitions, we considered the well-established iMod benchmark \cite{lopez2011imod} comprising a couple of tens of proteins with a wide variety of motions (hinge, shear, allosteric, and complex motions) often associated with ligand or partner binding. PETIMOT\emph{-5folds} predicted these transitions with high accuracy, achieving a 86\% success rate with an average Min. LS error of 0.41 $\pm$ 0.18 and an average Min. Magnitude error of 0.14 $\pm$ 0.07. For fold switches, we compiled a dataset of six metamorphic proteins from \cite{wayment2023predicting}. PETIMOT\emph{-5folds} achieved a success rate of 37\% on these challenging cases, with a Min. LS error of 0.67 $\pm$ 0.17 and a Min. Magnitude error of 0.25 $\pm$ 0.14. Our approach performed particularly well on KaiB, also highlighted in \cite{wayment2023predicting}. The Min. LS error is 0.45 starting from the ground state (PDB id: 2QKE, chain C) and 0.57 starting from the FS state (PDB id: 5JYT, chain A). 
Finally, we considered the ATPase NSF whose experimental structures correspond to ATP/ADP-bound states and 20S supercomplex conformations from cryo-EM studies \cite{zhao2015mechanistic,white2018structural}. The functionally relevant motions involve large-amplitude rigid-body domain movements and loop rearrangements. The first linear PCA mode explains 57\% of the variance and 4 modes are required to explain 90\%. PETIMOT successfully captures this complex motion subspace with Min. LS error as low as 0.32 and Global SS error of 0.30, demonstrating its ability to predict not just single motions but biologically meaningful motion subspaces. 

\begin{figure}[ht!]
    \centering
    \includegraphics[width=\linewidth]{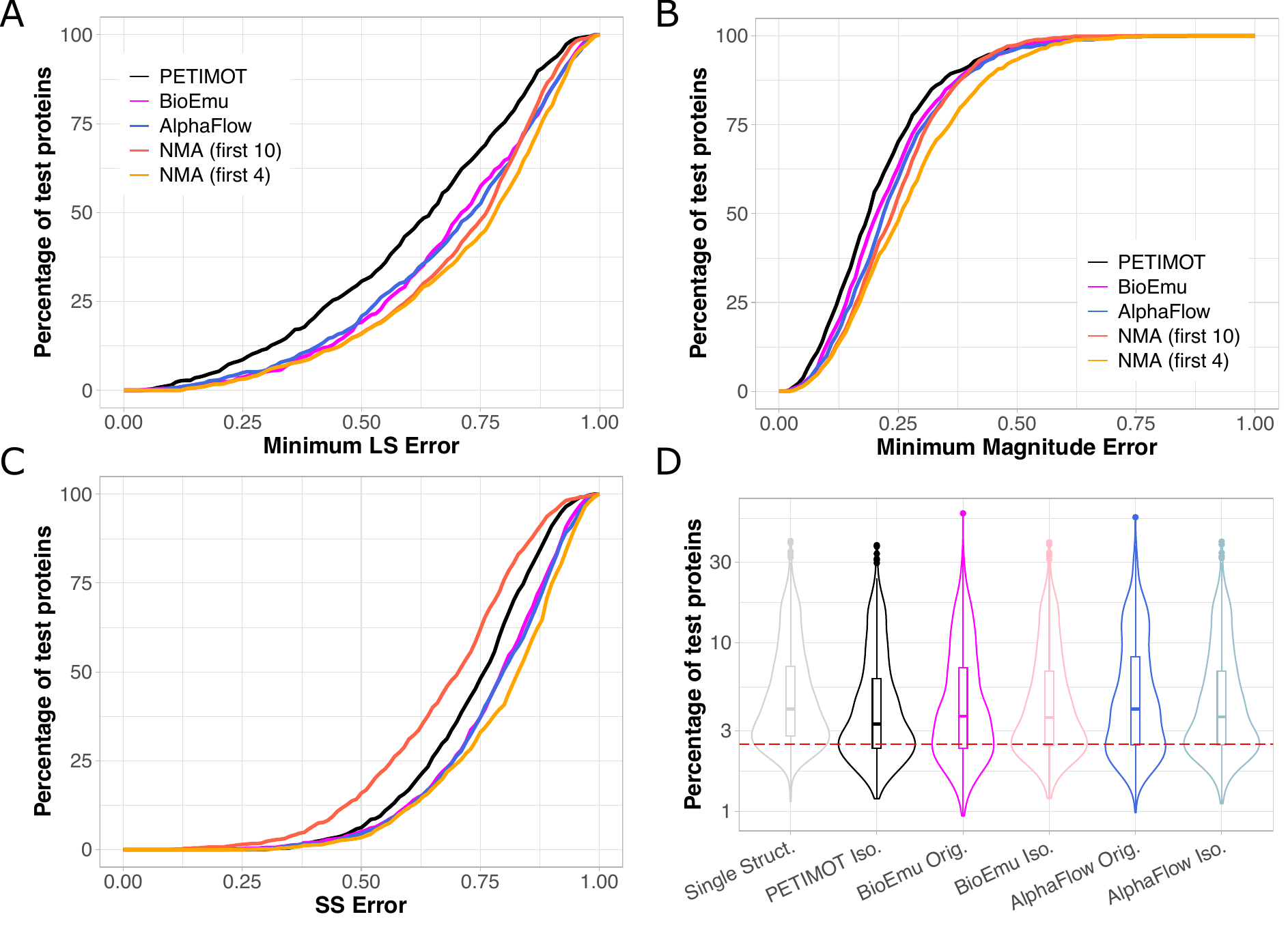}
    \caption{{{\bf PETIMOT's performance on the test set and comparison with other methods.}  PETIMOT\emph{-default} is compared with AlphaFlow, BioEmu, and the NMA on \emph{test824}. For each protein, we evaluate the 4 motions predicted by PETIMOT or inferred from AlphaFlow/BioEmu predicted ensembles against the 4 ground-truth motions. We allow the NMA more flexibility by considering the 10 lowest frequency modes. {\bf A.} Minimum LS Error, computed for the best matching pair of predicted and ground-truth motions. {\bf B.} Minimum Magnitude Error. {\bf C.} Global SS error, reflecting ground-truth subspace coverage. {\bf D.} Distributions of Max. Min. RMSD: within each generated ensemble, we identify the conformations closest to the reference  structures, and compute the maximum RMSD among them. If this value is below 2.5 \AA{} then we consider that all reference structures are covered by the ensemble. Single Struct. refers to the input structure given to PETIMOT. Orig. stands for the original ensembles generated directly from BioEmu and AlphaFlow while Iso. stands for isotropic ensembles generated from motions predicted by PETIMOT or inferred from BioEmu/AlphaFlow original ensembles.}}
    \label{fig:cumCurves}
\end{figure}
 
\paragraph{{Comparison with the physics-based unsupervised NMA.}}
We primarily compare PETIMOT with the NMA, a cost-effective approach for predicting the motion directions energetically accessible to a protein 3D structure. {We considered the ten normal modes with the lowest frequencies. This asymmetrical evaluation, compared to PETIMOT's four predicted motions, ensures a strong baseline and follows the suggestion that weighted mixtures of low-frequency modes can be more informative than individual modes \cite{kolossvary2024fresh}. PETIMOT produced acceptable predictions for almost 40\% of the dataset while the NMA's success rate is ~25\% (Fig. \ref{fig:indiv}D, Tables \ref{tab:fullComp} and \ref{tab:additional}), and PETIMOT achieved lower errors than the NMA in two thirds of the proteins (Fig. \ref{fig:cumCurves}A). While PETIMOT's individual predictions better match the ground-truth motions, the 10-mode subspace predicted by the NMA has higher overlap with the 4-mode ground-truth subspace than the 4-mode PETIMOT subspace (Fig. \ref{fig:cumCurves}C and Table \ref{tab:additional}, Global SS error of 0.67 $\pm$ 0.16 versus 0.73 $\pm$ 0.14). This suggests that low-frequency NMA modes collectively span the conformational subspace well — albeit at the cost of losing individual mode precision. When restricting to the first four normal modes, the Global SS error drastically increases to 0.79 $\pm$ 0.14 (Fig. \ref{fig:cumCurves}C). These results hold when varying the density and resolution of the Elastic Network Model (ENM, see Tables \ref{tab:fullComp} and \ref{tab:NMAsweeps}): among C$\alpha$ ENM variants, a cutoff of 10\AA{} yields the best performance, and switching to an all-atom representation provides only marginal gains (Success Rate 28.52\% versus 25.73\%), leaving PETIMOT's advantage intact across all configurations.} 
Moreover, PETIMOT {outperformed the NMA in terms of consistency across conformations (Fig. \ref{supfig:influRef}b) and} was 2.75 times faster at inference (Table \ref{tab:additional}).

\paragraph{{Comparison with generative models.}}
We considered the flow-matching or diffusion based generative models AlphaFlow and BioEmu as additional baselines. AlphaFlow was trained solely on the PDB while BioEmu was trained on massive amounts of experimental structures, 3D models, and MD conformations. To ensure fair comparison, we relied on both our motion-specific metrics and on commonly used metrics for comparing conformational ensembles (see Appendix \ref{evaluation}).
{We shall acknowledge a potential data leakage between AlphaFlow and BioEmu training data and our test set.}
PETIMOT outperforms both ensemble-based methods on motion subspace metrics {(Fig. \ref{fig:cumCurves}A-C)}, with a 43.57\% success rate versus $\sim$31\% for AlphaFlow and BioEmu {(Fig. \ref{fig:indiv}D)}, and a substantially lower Global SS Error {(Fig. \ref{fig:cumCurves}C and Table \ref{tab:additional}, 0.73 vs 0.77-0.78)}.
Cases where PETIMOT produces highly inaccurate predictions (Min. LS loss above 0.7) while the baselines are clearly successful (Min. LS loss below 0.4) are extremely limited (less than 5 per baseline, see for instance Fig. \ref{fig:2HCBC:arrow}).
See Appendix \ref{resultsA} for more details.
 
\begin{table}[bht!]
\caption{{\bf Comparison of PETIMOT's ability to generate conformational ensembles with respect to generative models.}  PETIMOT\emph{-default} is compared with AlphaFlow and BioEmu on \emph{test824}. RMS Fluctuation (RMSF) correlation was computed between conformational samples we generated from the ground-truth or predicted motions. The Min. RMS Deviations (RMSD) were computed between each of five experimental structures and the predicted ensembles {(see Appendix \ref{evaluation}). $^*$For AlphaFlow and BioEmu, the original ensemble produced by the method; for PETIMOT, the input structure with zero displacement.} Arrows indicate whether higher ($\uparrow$) or lower ($\downarrow$) values are better, best results highlighted in \textbf{bold}.}
\centering
\begin{tabular}{lccc}
\toprule
Metrics & PETIMOT & AlphaFlow & BioEmu \\
\toprule
{\it RMSF Correlation} $\uparrow$ & {\bf 0.59 $\pm$ 0.23} & 0.51 $\pm$ 0.25 & 0.52 $\pm$ 0.27 \\ 
\midrule
\multicolumn{4}{l}{{\it Coverage, Max. Min. RMSD $< $2.5\AA{} (\%)} $\uparrow$}\\
original ensemble$^*$ & 17.05 & 25.64 & {\bf 28.90} \\
isotropic sampling & {\bf 29.96} & 26.12 & 26.83\\
\multicolumn{4}{l}{{\it Avg Min. RMSD (\AA{})} $\downarrow$} \\
original ensemble$^*$ & 3.52 $\pm$ 3.04 & 4.77 $\pm$ 4.99 & {\bf 4.29 $\pm$ 4.34}  \\
isotropic sampling & {\bf 3.46 $\pm$ 2.83} & 3.54 $\pm$ 2.97 & 3.52 $\pm$ 2.95 \\
\bottomrule
\end{tabular}
\label{tab:ensembles}
\end{table}

Beyond predicting linear motions, PETIMOT allows straightforwardly generating conformational ensembles or trajectories by deforming the input protein 3D structure. We showcase this functionality on the xylanase A from {\it Bacillus subtilis} (Fig. \ref{fig:indiv}A-B). We used {PETIMOT's best predicted motion} to generate physically realistic conformations representing the open-to-closed transition of xylanase A thumb. More broadly, the {conformational ensembles generated by randomly sampling PETIMOT's predicted motions yield a RMSF Pearson correlation of 0.59 $\pm$ 0.23 against ensembles derived from ground-truth motions (Table \ref{tab:ensembles}), outperforming AlphaFlow (0.51 $\pm$ 0.25) and BioEmu (0.52 $\pm$ 0.287). When assessed against representative experimental structures, PETIMOT ensembles achieve higher or similar coverage compared to the baselines (Table \ref{tab:ensembles} and Fig. \ref{fig:cumCurves}D). All representatives are approximated with RMS deviations lower than 2.5\AA{} in 29.96\% of the cases. This is 13 percentage points higher than when considering only the input structure alone, which we used as a trivial lower bound for PETIMOT. By comparison, BioEmu's original ensembles cover 28.90\%, confirming its ability to produce diverse and biologically relevant conformations. The average minimum RMSD to experimental structures tells a consistent story, with PETIMOT's isotropic sampling achieving the lowest value (Table \ref{tab:ensembles}, 3.46 $\pm$ 2.83\AA{}). These results demonstrate that the quality of PETIMOT's predicted motion subspace — as measured by our specialized metrics — directly translates to competitive conformational flexibility and coverage, even when using a simple isotropic sampling protocol.}

\paragraph{Comparison of problem formulations.}
Our base model combining the LS and SS loses with equal weights outperforms all three individual losses, LS, SS, and LS (Fig. \ref{fig:extraPerfLosses} and \ref{fig:extraPerf2}). It strikes an excellent balance between approximating individual motions with high accuracy (Fig. \ref{fig:extraPerfLosses}a) and globally covering the motion subspaces (Fig. \ref{fig:extraPerfLosses}b). By comparison, the SS and IS losses tend to underperform on individual motions while the LS loss tends to provide lower coverage of the ground-truth subspaces. 
See Appendix \ref{ablation} for more details.

 \paragraph{Contribution of sequence and structure features.} 
 We performed an ablation study to assess the contribution of sequence and structure information to our architecture. Our results show that ProstT5 slightly outperforms the more recent and larger pLM, ESM-Cambrian 600M \cite{esm2024cambrian} (Fig. \ref{fig:ablation1}). Geometrical information about protein structure provides the most significant contribution, as replacing ProstT5 embeddings with random numbers has only a small impact on network performance. Conversely, the network's performance without structural information strongly depends on the chosen pLM. While the structure-aware embeddings from ProstT5 partially compensate for missing 3D structure information, relying solely on ESM-C embeddings results in poor performance (Fig. \ref{fig:ablation1}). Moreover, connecting each residue to its 15 nearest neighbours (sorted according to C$\alpha$-C$\alpha$ distances) in the protein graph results in lower performance compared to introducing randomly chosen edges or even fully relying on random connectivity (Fig. \ref{fig:abl:graph}).  
See Appendix \ref{ablation} for more details.

  \paragraph{Generalisation to MD data.} 
  To further assess PETIMOT's robustness, we evaluated it on MD trajectories from the ATLAS dataset \cite{vander2024atlas}. We identified 400 protein chains common to both the ATLAS set and our dataset, providing an independent MD benchmark (see Appendix \ref{atlas}). To ensure rigorous evaluation without data leakage, for each ATLAS protein chain we used the corresponding PETIMOT\emph{-5folds} model trained on the fold where that specific chain was held out from training (ensuring no training exposure). PETIMOT\emph{-5folds} achieved a 60\% success rate on this MD data -- with a Min. LS error $0.55 \pm 0.19$, Min. Magnitude error $0.17 \pm 0.11$, and Global SS error $0.60 \pm 0.16$. These performance metrics are significantly better than those obtained on experimental structures. Moreover, the association between Min. LS error and SS error is higher -- Adjusted R-squared of 0.71 versus 0.60 on the PDB dataset (Fig. \ref{fig:minLS_SS}). These results demonstrate that PETIMOT generalises to MD data without re-training nor fine-tuning.
  
\paragraph{Limitations.}

PETIMOT’s relatively modest success rate may be partially explained by incomplete and biased functional state sampling in the PDB, where predicted motions through evolutionary transfer may correspond to functionally relevant conformational states that have not been structurally resolved, and experimental artifacts ({\it e.g.}, of crystallographic origin, or due to sequence engineering). Our working hypothesis is that a part of the conformational manifold represents functionally relevant motions. To address this challenge, we designed our training loss function specifically to evaluate submanifolds by calculating the minimum error between each reference motion and the set of predicted motions, allowing the model to capture conformational diversity while mitigating the impact of potential artefacts. By comparison, the Atlas MD trajectories represent an easier case but they are limited to equilibrium distributions of monomeric proteins and do not account for conformational changes induced by partner or ligand binding.

{Furthermore, while PETIMOT's predicted motions support competitive conformational ensemble generation through simple isotropic sampling, generative models such as BioEmu and AlphaFlow offer complementary advantages: trained end-to-end to produce diverse ensembles, they may better capture rare or large-amplitude conformational states, at the cost of significantly higher computational requirements and reduced interpretability.}

In addition, our approach is limited to modeling protein motions as linear displacement vectors. While this approximation is sufficient to describe most of the observed conformational heterogeneity, it remains inadequate for modeling highly complex non-linear deformations. Furthermore, deforming proteins structures along linear motion direction may produce unrealistic conformations at large amplitudes.  A possible solution yet to be investigated would be nonlinear extrapolation techniques
 widely used in molecular mechanics \cite{lopez2011imod,hoffmann2017nolb}. 

\section{Conclusion}
In this work, we have proposed a new perspective on the problem of capturing protein continuous conformational heterogeneity. Our approach directly infers compact and continuous representations of protein motions. Our comprehensive analysis of PETIMOT's predictive capabilities demonstrates its performance and utility for understanding how proteins deform to perform their functions. It shows that accurate motion subspace prediction—PETIMOT's core strength—provides a strong foundation for modeling protein functional dynamics, while offering interpretability and efficiency advantages over generative models for conformational sampling. PETIMOT-generated structures, while not being accurate in a thermodynamic sense, can help practitioners quickly assess possible dynamics or seed other workflows like heterogeneous cryo-EM reconstruction. Our work opens ways to future developments in protein motion manifold learning, with exciting potential applications in protein engineering and drug development.
%


\clearpage   
\appendix 

\counterwithin{figure}{section}
\counterwithin{table}{section}
\counterwithin{equation}{section}
\counterwithin{algorithm}{section}

\renewcommand{\thefigure}{\Alph{section}.\arabic{figure}}
\renewcommand{\thetable}{\Alph{section}.\arabic{table}}
\renewcommand{\theequation}{\Alph{section}.\arabic{equation}}
\renewcommand{\thealgorithm}{\Alph{section}.\arabic{algorithm}}


\section*{Appendices}

\section{Invariance of the proposed losses}
\label{app:loss:invariance}

\begin{theorem}
SS Loss is invariant under unitary transformations of $X$ and $Y$ subspaces.
\end{theorem}

\begin{proof}
Without loss of generality, let us assume that we apply a unitary transformation $U \in \mathbb R^{K \times K}$ to a subspace $X^\perp \in \mathbb R^{3N \times K}$, such that 
the result $X' = X^\perp U$, with $X' \in \mathbb R^{3N \times K}$, spans the same subspace as $X^\perp$, as it is a linear combination of the original basis vectors from $X^\perp$.
 Then, let us rewrite the SS loss as 
\begin{equation}
 \text{SS Loss} = 1 - \frac{1}{\min(K,L)} \sum_{k=1}^{K} \sum_{l=1}^{L} ( {\mathbf y}_{k}^T W^{1\over 2} {\mathbf x^{\perp}_{l}} )^2 
 = 1- \frac{1}{\min(K,L)} || Y^T  W^{1\over 2} X^\perp||^2_F.
\end{equation} 
As the Frobenius matrix norm is invariant under orthogonal, or more generally, unitary, transformations, $|| Y^T  W^{1\over 2} X^\perp U||^2_F = || Y^T  W^{1\over 2} X^\perp||^2_F$, which completes the proof.
\end{proof}

\begin{corollary}
The SS loss is invariant to the direction permutations in the Gram-Schmidt orthogonalization process.
\end{corollary}
\begin{proof}
Let us consider two linear subspaces  $X^\perp_1$ and  $X^\perp_2$ resulting from the  Gram-Schmidt orthogonalization of  $X$, where we arbitrarily choose the order of the orthogonalization vectors. Both  $X^\perp_1$ and  $X^\perp_2$ will span the same subspace as $X$,
and since  both $X^\perp_1$ and $X^\perp_2$ are also orthogonal, one is a unitary transformation of the other, $X^\perp_2 = X^\perp_1 U$, which completes the proof.
\end{proof}

\begin{theorem}
IS Loss is invariant under unitary transformations of $X$ and $Y$ subspaces.
\end{theorem}

\begin{proof}
Following the previous proof, without loss of generality, let us assume that we apply an orthogonal (unitary) transformation $U \in \mathbb R^{K \times K}$ to a subspace $X \in \mathbb R^{3N \times K}$, such that 
the result $X' = X U$, with $X' \in \mathbb R^{3N \times K}$, spans the same subspace as $X$.
 Then, let us rewrite the IS loss as 
\begin{equation}
\begin{split}
 \text{IS Loss} &=  
 \frac{1}{L^2} \sum_{l=1}^{L} \sum_{m=1}^{L}  ( \textbf{x}_{l}^T W \textbf{x}_{m} )^2
 - 
 \frac{1}{L\min(K,L)} \sum_{k=1}^{K} \sum_{l=1}^{L} ( \textbf{y}_{k}^T W^{\frac{1}{2}} \textbf{x}_{l} )^2 \\
 &=  \frac{1}{L^2} ||X^T  W X ||^2_F
 - \frac{1}{L\min(K,L)}  || Y^T  W^{1\over 2} X||^2_F.
\end{split}
\end{equation}
As the Frobenius matrix norm is invariant under orthogonal transformations, 
$|| Y^T  W^{1\over 2} X U||^2_F = || Y^T  W^{1\over 2} X||^2_F$, 
and
$ ||U^T X^T  W X U||^2_F =  ||X^T  W X ||^2_F$,
which completes the proof.
\end{proof}



\section{Methods details}
\label{methods_details}

\subsection{Training data}
\label{trainingA}
\paragraph{Conformational collections.} To generate the training data, we used DANCE \cite{lombard2024explaining} to construct a non-redundant set of conformational collections representing the entire PDB as of June 2023. Wherever possible, we enhanced the data quality by replacing raw PDB coordinates with their updated and optimized counterparts from PDB-REDO \cite{joosten2014pdb_redo}. Each conformational collection was designed to include only closely related homologs, ensuring that any two protein chains within the same collection shared at least 80\% sequence identity and coverage. Collections with insufficient data points were excluded as we require at least 5 conformations. To simplify the data, we retained only C$\alpha$ atoms (option \texttt{-c}) and accounted for coordinate uncertainty by applying weights (option \texttt{-w}).
{To reduce structural redundancy of crystallographic artifacts and non-biological conformations, we removed any conformation A deviating by less than 0.1 \AA~ from another one B, provided that the sequence of A is identical to or included in that of B.}

\paragraph{Handling missing data.}
The conformations in a collection may have different lengths reflected by the introduction of gaps when aligning the amino acid sequences. {As detailed in \cite{lombard2024explaining}, we} fill these gaps with the coordinates of the conformation used to center the data. {The imputed positions have zero centered coordinates and therefore contribute no variance to the decomposition — making this a fairly neutral imputation strategy.} Moreover, to explicitly account for data uncertainty, we assign confidence scores to the residues and include them in the structural alignment step and the eigen decomposition. 
The confidence score of a position $i$ reflects its coverage in the alignment, 
\begin{equation}
    w_i = \frac{1}{m}
    \sum_S 
    \mathds{1}_{a_i^S \ne \text{"X"}}, 
    \label{eq:coverage}
\end{equation}

where "X" is the symbol used for gaps and $m$ is the number of conformations. The structural alignment of the $j$th conformation onto the reference conformation amounts to determining the optimal rotation that minimises the following function
\cite{Kabsch:a12999,kearsley1989orthogonal},
\begin{equation}
E = \frac{1}{\sum_i w_i} \sum_i w_i (r^c_{ij} - r^c_{i0})^2,
\end{equation}
where $r^c_{ij}$ is the $i$th centred coordinate of the $j$th conformation and $r^c_{i0}$ is the $i$th centred coordinate of the reference conformation. The resulting aligned coordinates are then multiplied by the confidence scores prior to the PCA, as we explain below.
{The weighting scheme effectively prevents large deviations in uncertain regions, typically highly localised loop motions, from dominating the variance \cite{lombard2024explaining}.}

\paragraph{Eigenspaces of positional covariance matrices.} The Cartesian coordinates of each conformational ensemble can be stored in a matrix $R$ of dimension $3N \times m$, where $N$ is the number of residues (or positions in the associated multiple sequence alignment) and {$m$} is the number of conformations. Each position is represented by a C-$\alpha$ atom. We compute the coverage-weighted (to account for missing data, as explained above) covariance matrix as in Eq. \ref{eq:covariance}.
The covariance matrix is a $3N\times 3N$ square matrix, symmetric and real.  

We decompose $C$ as $C=VDV^T$, where $V$ is a $3N\times 3N$ matrix with each column defining a sqrt-coverage-weighted eigenvector or a principal component that we interpret as a {\em linear motion}. 
$D$ is a diagonal matrix containing the eigenvalues. 
Specifically, the $k$th principal component was expressed as a set of 3D (sqrt-coverage-weighted)
displacement vectors ${\vec x}^{GT}_{ik}, i=1, 2, ...L$ for the $L$ C$\alpha$ atoms of the protein residues. To enable cross-protein comparisons, the vectors were normalized such that $\sum{i=1}^L | \vec{x}^{GT}_{ik}|^2 = L$. The sum of the eigenvalues $\sum_{k=1}^{3m}\lambda_k$ amounts to the total positional variance of the ensemble (measured in \AA$^2$) and each eigenvalue reflects the amount of variance explained by the associated eigenvector.

\paragraph{Data augmentation.} The reference conformation used to align and center the 3D coordinates corresponds to the protein chain with the most representative amino acid sequence. To increase data diversity, four additional reference conformations were defined for each collection. At each iteration, the new reference conformation was selected as the one with the highest RMSD relative to the previous reference. This iterative strategy maximizes the variability of the extracted motions by emphasizing the impact of changing the reference.

\subsection{Message passing}
\label{MPBlockA}

The node embeddings and predicted motion vectors are updated iteratively according to the following Algorithm \ref{alg:PETIMOT} 
{, where
MessageMLP stands for one hidden layer of size 256, followed by a GELU activation and a dropout layer with rate 0.4.}
\begin{algorithm}[H]
\caption{PETIMOT Message Passing Block}
\label{alg:PETIMOT}
\begin{algorithmic}[1]
\Function{messagePassing}{$\{\mathbf{s}_i\},\{\vec{x}_i\},\{\mathcal{N}eigh(i)\},\{R_{ij}, e_{ij}\}$}:\\
\# $\{\mathbf{s}_i\}_{i=1}^N$   \Comment Node embeddings\\
\# $\{\vec{x}_i\}_{i=1}^N$   \Comment Motion vectors in local frames\\
\# $\{\mathcal{N}eigh(i)\}_{i=1}^N$   \Comment Node neighborhoods\\
\# $\{R_{ij}, e_{ij}\}$   \Comment Relative geometric features
\For{$i = 1$ to $N$}
    \For{$j \in \mathcal{N}eigh(i)$}
         \State $\vec{x}_j^i \gets R_{ij}\vec{x}_j$   \Comment Project motion in frame $i$
         \State $m_{ij} \gets \text{MessageMLP}(\mathbf{s}_i, \mathbf{s}_j, \vec{x}_i, \vec{x}_j^i, e_{ij})$
    \EndFor
    \State $m_i \gets \text{Mean}_j(m_{ij})$   \Comment Aggregate messages
    \State $\mathbf{s}_i \gets \mathbf{s}_i + \text{LayerNorm}(m_i)$   \Comment Update embedding
     \State $\vec{x}_i \gets \vec{x}_i + \text{Linear}([\mathbf{s}_i, \vec{x}_i])$   \Comment Update motion
\EndFor
\State \Return $\{\mathbf{s}_i\}_{i=1}^N, \{\vec{x}_i\}_{i=1}^N$
\EndFunction
\end{algorithmic}
\end{algorithm}

\subsection{SE(3)-equivariant features}
\label{featuresA}
We represent protein structures as attributed graphs. The node embeddings are computed with the pre-trained protein language model ProstT5 \cite{heinzinger2023prostt5}. It is a fine-tuned version of the sequence-only model T5 that translates amino acid sequences into sequences of discrete structural states and reciprocally.

The edge embeddings are computed using SE(3)-invariant features derived from the input backbone, similarly to prior works \cite{ingraham2023illuminating,dauparas2022robust,ingraham2019generative}. Specifically, the features associated with the edge $e_{ij}$ from node (atom) $i$ to node (atom) $j$ are:
\begin{itemize}
    \item {\bf Quaternion representation:} A 4-dimensional quaternion encoding the relative rotation $R_{ij}$ between the local reference frames of residues $i$ and $j$.
    
    \item {\bf Relative translation:} A 3-dimensional vector representing the translation $\vec{t}_{ij}$ between the local reference frames.
    
    \item {\bf Chain separation:} The sequence separation between residues $i$ and $j$, encoded as $\log(|i-j| + 1)$.
    
    \item {\bf Spatial separation:} The logarithm of the Euclidean distance between residues $i$ and $j$, computed as $\log(\|\vec{t}_{ij}\| + \epsilon)$, where $\epsilon = 10^{-8}$.
    
    \item {\bf Backbone atoms distances:} Distances between all backbone atoms (N, C$\alpha$, C, O) at residues $i$ and $j$, encoded through a radial basis expansion. For each pairwise distance $d_{ab}$, we compute:
    \begin{equation}
        f_k(d_{ab}) = \exp\left(-\frac{(d_{ab} - \mu_k)^2}{2\sigma^2}\right),
    \end{equation}
    where $\{\mu_k\}_{k=1}^{20}$ are centers spaced linearly in $[0,20]$ \AA~and $\sigma=1$ \AA. This creates a 16 $\times$ 20 = 320 dimensional feature vector, as we have 16 pairwise distances (4 $\times$ 4 atoms) each expanded in 20 basis functions.
\end{itemize}

\subsection{Training procedure} 
\label{ref:traning}

For the \emph{default} version, we randomly split the 7,335 conformational collections defined with DANCE into training, validation, and test sets with a ratio 70:15:15. The data augmentation procedure resulted in 5,119 $\times$ 5 = 25,595 training samples and 1,099 $\times$ 5 = 5,495 validation samples.

For the \emph{stringent} version, we considered clusters of protein chains defined at 30\% sequence identity and 80\% sequence coverage using MMseqs2 \cite{Steinegger2017}. These clusters define distant protein families and we refer to them as clus-30 in the following. We kept the test proteins used PETIMOT-\emph{default} as is and we removed all collections belonging to the same clus-30 clusters as these proteins from the training and validation sets. Then, we re-defined a training-validation random split at the level of the clus-30 clusters with a 9:1 ratio. This operation ensures that any pair of training-validation, training-test or validation-test collections do not share more than 30\% sequence identity. Finally, for each training or validation clus-30 cluster, we randomly drew 5 samples. This step ensures that each protein family is evenly represented in the training and validation sets. We also redundancy reduced \emph{test824}, keeping only one protein for each clus-30 cluster, which led to 734 proteins (\emph{test734}).

For the \emph{5fold} version, we conducted a 5-fold cross-validation experiment with strict similarity filtering over the full training set comprising 36,675 samples. We ensured  a two-stage similarity filtering process:

\begin{enumerate}
\item \textbf{Structural similarity removal:} First, we used FoldSeek \cite{van2024fast} to cluster protein chains using an e-value threshold of 1e-2. Any two chains belonging to two different clusters do not share  significant structural similarity.
\item \textbf{Sequential similarity removal:} We randomly partitioned the dataset into 5 folds and applied a cross-validation procedure. We implemented an additional sequence similarity-based filtering using MMseqs2. Specifically, for each fold, we removed from the training set (80\% of the data) the protein chains sharing more than 30\% sequence identity with any of the chains from the test set (20\%).
\end{enumerate}


The models were optimized using AdamW \cite{loshchilov2018decoupled} with a learning rate of 5e-4 and weight decay of 0.01. We employed gradient clipping with a maximum norm of 10.0 and mixed precision training with PyTorch's Automatic Mixed Precision. The learning rate was adjusted using torch's ReduceLROnPlateau scheduler, which monitored the validation loss, reducing the learning rate by a factor of 0.2 after 10 epochs without improvement. Training was performed with a batch size of 32 for both training and validation sets. We implemented early stopping with a patience of 50 epochs, monitoring the validation loss. The model achieving the best validation performance was selected for final evaluation.
We trained the model on a single NVIDIA A100-SXM4-80GB GPU. One epoch took about 9 minutes of real time.

\subsection{Evaluation and comparison with other methods
\label{evaluation}}

We primarily evaluated PETIMOT on a test set of 1~117 proteins, reduced to 824 (\emph{test824}) to comply with other methods' requirements (see below). In addition, our 5-fold cross-validation training procedure allowed us to evaluate PETIMOT predictive capacity on the full dataset of 36,675 samples and systematically compare it with the NMA (Table \ref{tab:fullComp}).

\subsubsection{Evaluation protocol and metrics}

We primarily relied on three metrics defined in the main text to evaluate PETIMOT and compare it with other methods: the LS error, the magnitude error, and the Global SS error. We computed the LS error and the magnitude error either on the best-matching pair of predicted and ground-truth motions (Min.) or on the full set of matched pairs determined through optimal linear assignment (OLA). In addition, we included several commonly used metrics for assessing conformational ensembles: RMSF correlation and minimum RMSD to experimental structures. 

To generate conformational ensembles, we deformed each test protein's 3D coordinates along PETIMOT's four predicted motions. {Since we do not have access to the oracle eigenvalues at inference, we sampled displacement 
coefficients uniformly on a hypersphere of fixed radius. Formally, given 
$K$ predicted modes assembled in the matrix $\mathbf{X} \in \mathbb{R}^{3N \times K}$, 
a conformational sample is generated as:
\begin{equation}
\mathbf{r} = \mathbf{r}_0 + \mathbf{X}\mathbf{c}, \quad 
\mathbf{c} = \frac{\mathbf{z}}{\|\mathbf{z}\|} \cdot \sqrt{E_\text{total}}, \quad 
\mathbf{z} \sim \mathcal{N}(\mathbf{0}, \mathbf{I}_K)
\end{equation}
where $\mathbf{r}_0 \in \mathbb{R}^{3N}$ are the C$\alpha$ coordinates 
of the input reference structure, and $E_\text{total} = \epsilon \cdot N$ is a 
total energy budget scaling linearly with the number of residues $N$. This formulation ensures that all modes contribute equally in expectation, without imposing any ordering or relative weighting among predicted modes. We set $\epsilon = 1$ in all experiments, giving a displacement scale of $\sqrt{N}$ \AA{}. This choice is motivated by the substantial range of conformational diversity observed in the experimental collections of our benchmark, with a mean RMSD of 3.70 $\pm$ 3.36 \AA{} between conformations within each ensemble and a median maximum RMSD of 3.73 \AA{} across ensembles (up to 40.21 \AA{}).}

For RMSF correlation, we {additionally generated ensembles from ground-truth motions and corresponding eigenvalues. We randomly sampled deformation amplitudes from a Gaussian distribution with variance proportional to each ground-truth motion's eigenvalue. We} then compared per-residue fluctuations between predicted and ground-truth ensembles. For experimental structure coverage, we leveraged our iterative strategy for data augmentation (Appendix \ref{trainingA}) to select five diverse conformations from the ground-truth test collections. Then, for each experimental structure, we computed its Min. RMSD to the closest conformation in the predicted motion-derived ensemble. We defined coverage as the fraction of test proteins for which all five structures have a Min. RMSD below 2.5 \AA{}, a threshold commonly used for assessing conformational similarity in C-$\alpha$-only comparisons.

\subsubsection{Comparison with the Normal Mode Analysis} 

We compared our approach with the physics-based unsupervised Normal Mode Analysis (NMA) method \cite{hayward1995collective}. The NMA takes as input a protein 3D structure and builds an elastic network model where the nodes represent the atoms and the edges represent springs linking atoms located close to each other in 3D space. The four lowest normal modes are obtained by diagonalising the mass-weighted Hessian matrix of the potential energy of this network. We used the highly efficient NOLB method, version 1.9, downloaded from \url{https://team.inria.fr/nano-d/software/nolb-normal-modes/} \cite{hoffmann2017nolb} to extract the first $K$ normal modes from the test protein 3D conformations. Specifically, we used the following command {by default},
\begin{verbatim}
    NOLB INPUT.pdb -c 10 -x -n 10 --linear -s 0 --format 1 --hetatm
\end{verbatim}
{where the elastic network is defined using a distance cutoff (option -c) of 10 \AA{} and the input PDB file contains only C$\alpha$ atoms. Additionally, we tested the impact of varying the cutoffs (7.5 \AA{}, 13\AA{}) and of considering all atoms, using SCWRL4 \cite{krivov2009improved} to reconstruct the side-chains trimmed during data pre-processing.} We evaluated the NMA using exactly the same metrics and protocols as those used for PETIMOT, since both methods output a set of linear motions. {While we retained only the first four NMA modes for our primary evaluation, we additionally explored the impact of including all ten first modes when computing the evaluation metrics.}

\subsubsection{Comparison with generative models}

We considered the flow-matching based framework AlphaFlow and the diffusion-based Biomolecular Emulator (BioEmu) as additional baselines.

\paragraph{Conformational ensemble generation with AlphaFlow.}
Out of a total of 1~117 proteins comprised in our test set, we excluded 293 proteins because they were too long ($>$450 amino acids) to be handled by AlphaFlow in a reasonable amount of time using our computing resources. We downloaded the distilled "PDB" models from \url{https://github.com/bjing2016/alphaflow}. 
We executed AlphaFlow using the following command,
\begin{verbatim}
python predict.py --noisy_first --no_diffusion --mode alphafold 
--input_csv seqs.csv --msa_dir msa_dir/ 
--weights alphaflow_pdb_distilled_202402.pt --samples 50 
--outpdb output_pdb/
\end{verbatim}
AlphaFlow relies on OpenFold \cite{ahdritz2024openfold} to retrieve the input multiple sequence alignment (MSA). 
We used AlphaFlow to generate 50 conformations for each test protein.

\paragraph{Conformational ensemble generation with BioEmu.}

We ran BioEmu through its python API by reading the input FASTA file then launching the \emph{sample} function as
\begin{verbatim}
sample(sequence=record.seq, num_samples=50,
output_dir=f'./{record.id}')
\end{verbatim}
BioEmu generated up to 50 conformations per test protein, with an average of 44 $\pm$ 8 conformations.

\paragraph{Motion and conformation comparisons.}

While PETIMOT predicts a set of linear motions from an input protein sequence and structure, AlphaFlow and BioEmu predict a conformational ensemble from an input protein sequence. As a consequence, the input structure given to PETIMOT might not lie within the conformational ensemble predicted the generative models and our evaluation framework needs adaptation to ensure fair comparison. Firstly, to directly compare motions with our three main metrics, we aligned all members of the AlphaFlow or BioEmu original ensembles to the test protein conformations, with identity coverage weights, and extracted the principal linear motions. Secondly, to compare conformational ensembles using RMSF and RMSD metrics, we considered the original ensembles outputted by AlphaFlow or BioEmu, and, in addition, we generated new ensembles by deforming each test protein along the previously extracted principal linear motions. For this, we used the same "isotropic" sampling protocol as that used for PETIMOT conformational ensemble generation. 

We shall additionally mention that we did not filter or adapt our test set to the generative models. As a consequence, there can be data leakage between AlphaFlow and BioEmu train data and our test examples.
 
\subsubsection{Running times}

The running times were measured on an Intel(R) Xeon(R) W-2245 CPU @ 3.90GHz equipped with GeForce RTX 3090 for PETIMOT, AlphaFlow, and the NMA, and on an AMD Ryzen 9 7950X 16-Core CPU @ 5.88GHz equipped with NVIDIA RTX A6000 for BioEmu.
{We shall note that in deep architectures RTX A6000 can be up to 1.3 times faster compared to RTX 3090.}
 
\subsection{Ablation Studies \label{ablation}}
To understand the impact of different components on the performance of our model, we carried out ablation studies.  We list them below.

\begin{figure}[bht!]
    \centering
    \includegraphics[width=1\linewidth]{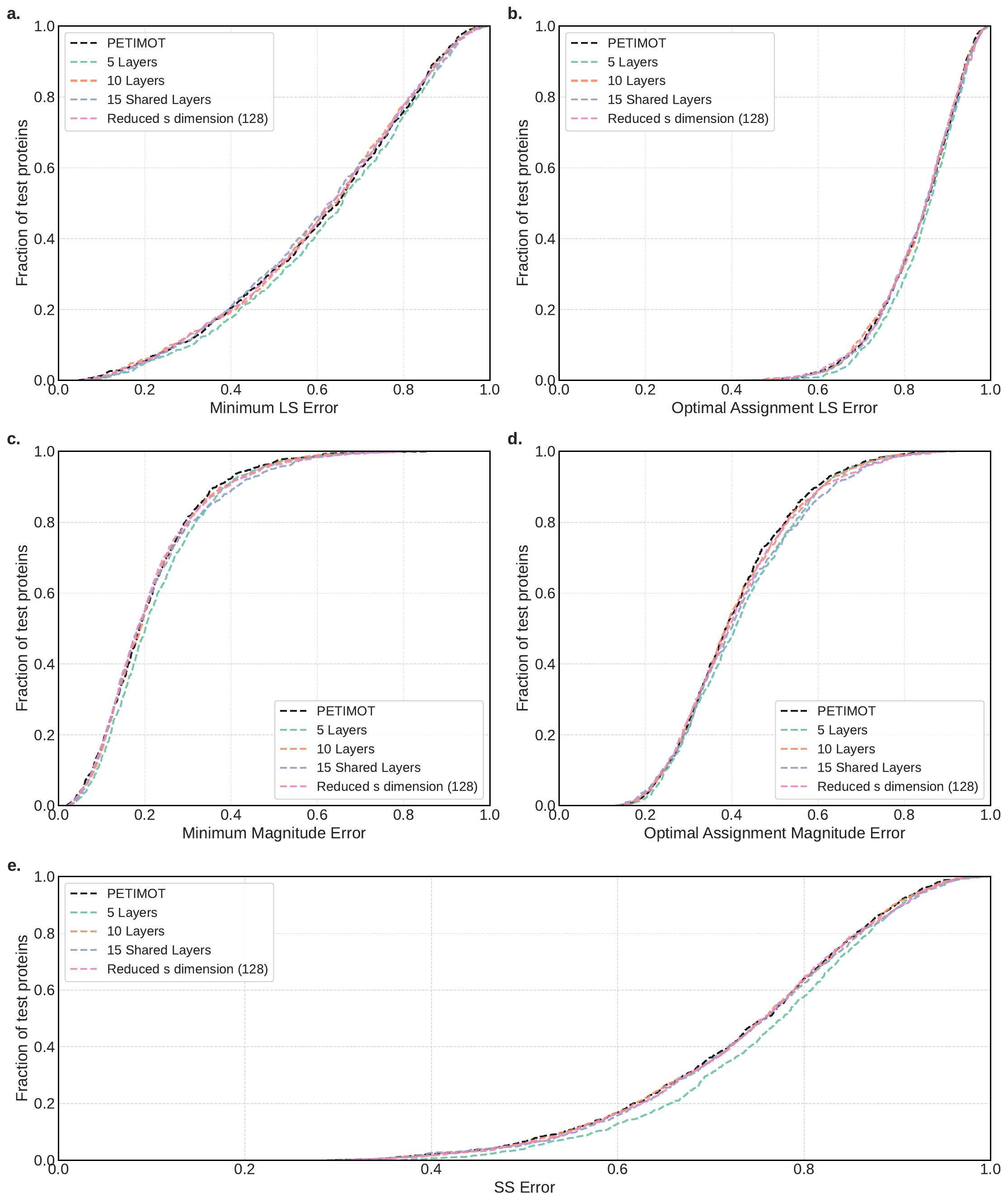}
    \caption{{\bf Network depth ablation.} We report cumulative curves for LS error (a-b), magnitude error (c-d), and SS error (e). For each protein, we computed the error either for the best-matching pair of predicted and ground-truth vectors (a,c) or for the best combination of four pairs of predicted and ground-truth vectors (b,d). We vary the number of layers in the network and the embedding dimension.}
    \label{fig:ablation3}
\end{figure}

\paragraph{Model architecture variations.}
\begin{itemize}
\item Network depth: We experimented with different numbers of message-passing layers (5 and 10 layers compared to our default value of 15 layers).
\item Layer sharing: We tested a variant where all message-passing layers share the same parameters, as opposed to our default where each layer has unique parameters.
\item Reduced internal embedding dimension: We tested a model with a smaller internal embedding dimension of 128 instead of the default 256.
\end{itemize}
Figure \ref{fig:ablation3} shows the evaluation of these modifications. A shallow 5-layers network underperforms on all evaluation metrics. The difference between other variants is not very significant.

\paragraph{Structure and sequence information ablation.}
\begin{itemize}
\item Structure ablation: We removed all structural information from the model to assess the importance of geometric features and the performance with the PLM embeddings only. We did it by removing the edge attributes of the input of the message passing MLP.
\item Sequence ablation: We ablated sequence information by replacing protein language model embeddings with random embeddings, testing them both with and without structural information.
\item Embedding variants: We evaluated a different protein language model (ESMC-600M), both with and without structural tokens.
\end{itemize}

\begin{figure}[th!]
    \centering
    \includegraphics[width=0.8\linewidth]{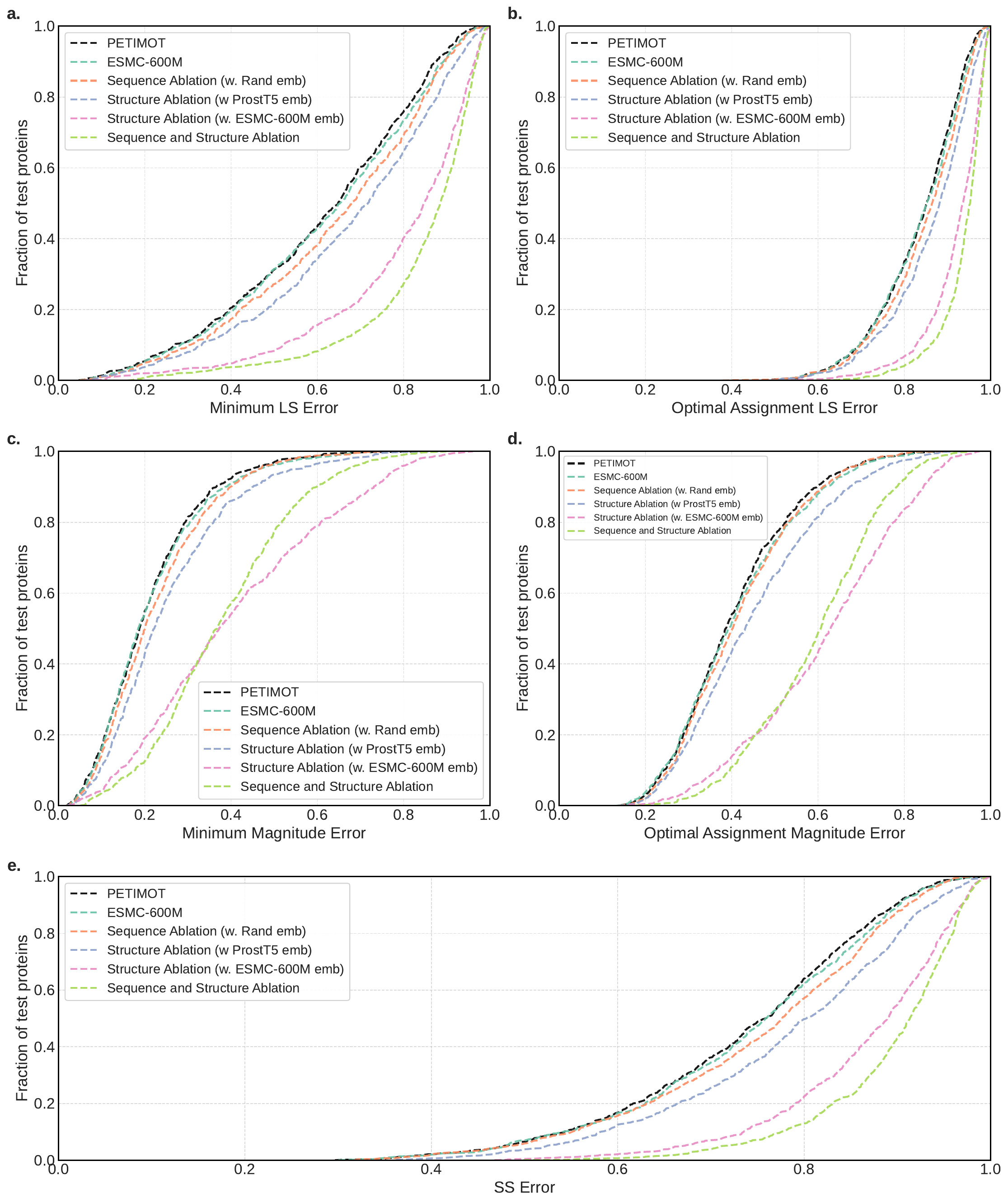}
    \caption{{\bf Structure and sequence information ablation study.} We report cumulative curves for LS error (a-b), magnitude error (c-d), and SS error (e). For each protein, we computed the LS and magnitude errors either for the best-matching pair of predicted and ground-truth vectors (a,c) or for the best combination of four pairs of predicted and ground-truth vectors (b,d).}
    \label{fig:ablation1}
\end{figure}

The evaluation results are shown in Fig. \ref{fig:ablation1}. The results demonstrate that while both ProstT5 and ESM-Cambrian 600M perform similarly when combined with structural information, removing structural features leads to markedly different outcomes. ProstT5 embeddings partially compensate for the missing structural information, likely due to their structure-aware training, while relying solely on ESM-C embeddings results in poor performance.

\paragraph{Problem formulation ablation.}
We analyzed different combinations of our loss terms (compared to our default balanced weights of  LS + SS):
\begin{itemize}
\item Least Square loss (LS): Using only the LS loss (weight 1.0).
\item Squared Sinus loss (SS): Using only the SS loss (weight 1.0).
\item Independent Subspaces (IS): Using only the IS loss (weight 1.0).
\end{itemize}

\begin{figure}[ht!]
     \centering
      \includegraphics[width=0.9\linewidth]{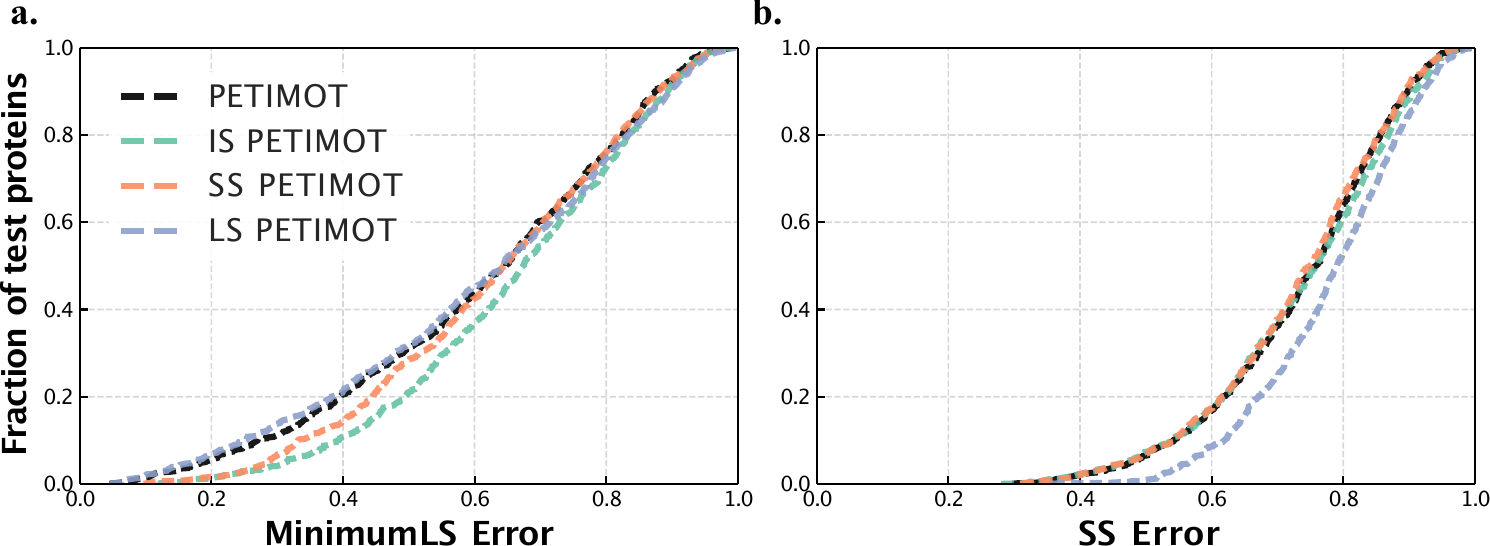}
\caption{{\bf Performance comparison of different problem formulations.} We report the cumulative curve for the minimum LS error (a, best matching pair) and the Global SS error (b) computed over the test set. The loss of the base model is LS + SS.}
     \label{fig:extraPerfLosses}
 \end{figure}

\begin{figure}[th!]
    \centering
    \includegraphics[width=\linewidth]{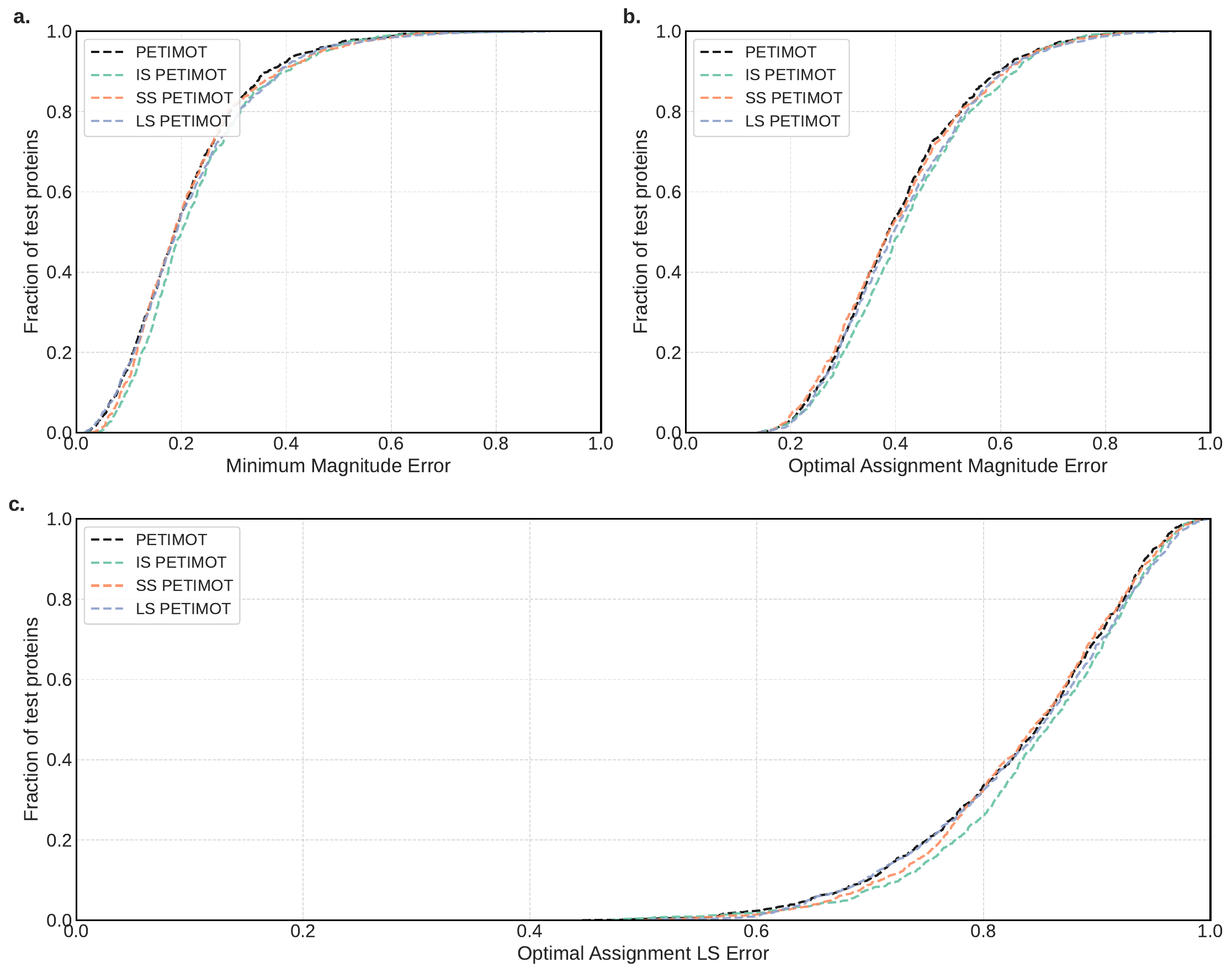}
    \caption{{\bf Performance comparison of different problem formulations.} We report cumulative curves for magnitude error, corresponding to the best-matching pair of predicted and ground-truth vectors (a, Min.) or the best combination of four pairs of predicted and ground-truth vectors (b, OLA) and OLA LS error (c). }
    \label{fig:extraPerf2}
\end{figure}

Figures \ref{fig:extraPerfLosses} and \ref{fig:extraPerf2} compare three individual losses with the default option. The IS problem formulation underperforms on all the metrics but the Global SS error. The default LS + SS formulation performs slightly better than those with individual loss components.

\begin{figure}[th]
    \centering
    \includegraphics[width=\linewidth]{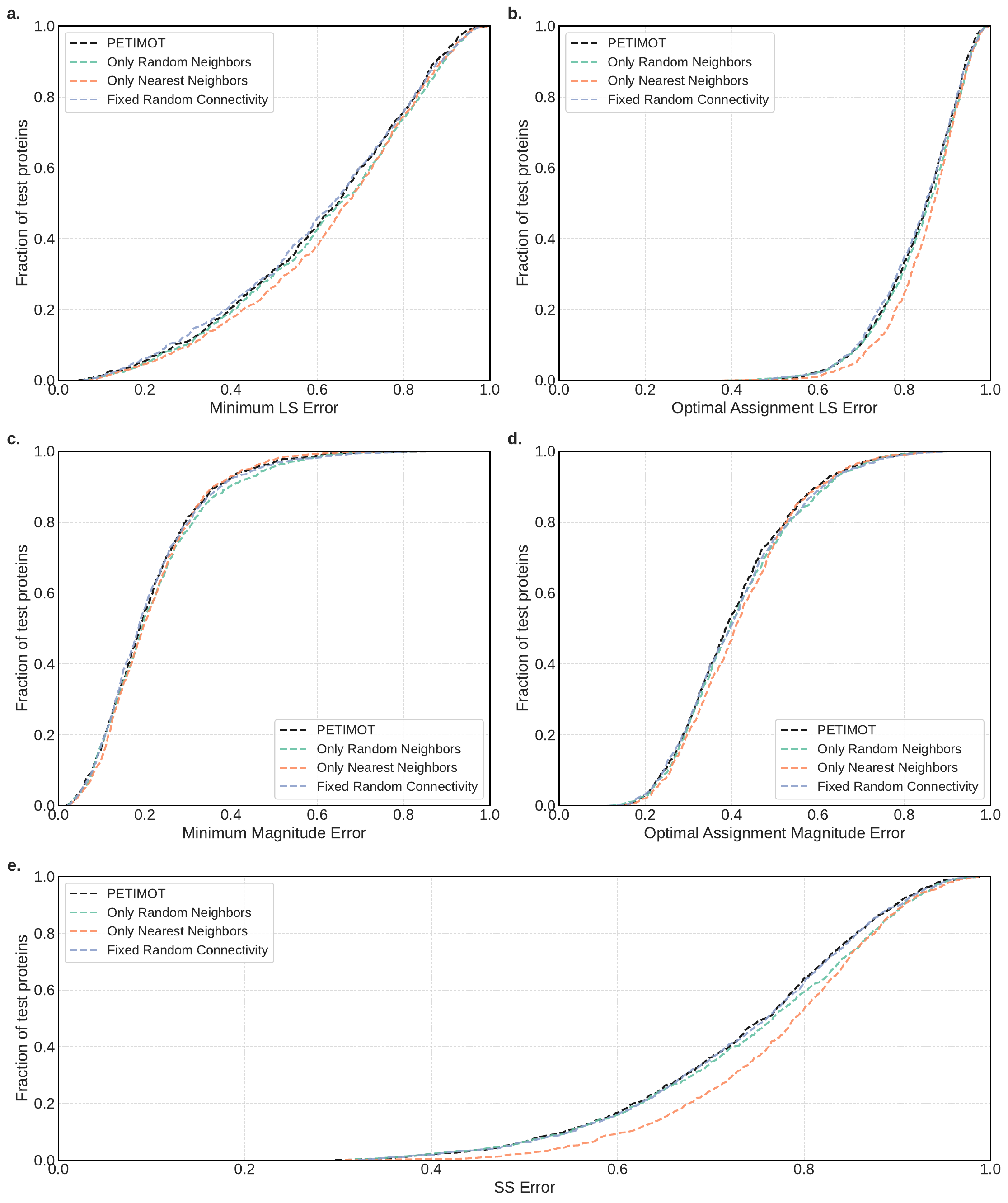}
    \caption{{\bf Graph connectivity ablation.} We report cumulative curves for LS error (a-b), magnitude error (c-d), and SS error (e). For each protein, we computed the error either for the best-matching pair of predicted and ground-truth vectors (a,c) or for the best combination of four pairs of predicted and ground-truth vectors (b,d). Only Random Neighbors: each residue (node) is connected to 15 randomly chosen residues and the connectivity changes after each layer. Only Nearest Neighbors: each residue (node) is connected to its 15 nearest neighbors in the input 3D structure. Fixed Random Connectivity: each residue (node) is connected to 15 residues randomly chosen at the beginning. }
    \label{fig:abl:graph}
\end{figure}

\paragraph{Graph connectivity ablation.}
We investigated different approaches to constructing the protein graph:
\begin{itemize}
\item Nearest neighbor-only: Using 15 nearest neighbors (sorted according to the corresponding C$\alpha$-C$\alpha$ distances) without random edges.
\item Random connections-only: Using 15 random edges without nearest neighbors. This set is updated between every layer at each epoch.
\item Static connectivity: Using a fixed set of random neighbors between the layers. This set is updated at each epoch.
\end{itemize}
Figure \ref{fig:abl:graph} shows the ablation results. We can see that the nearest neighbor-only setup underperforms on all the metrics. Among other options, the random connectivity-only option gets lower results at higher metrics values. The default option performs on par with the static connectivity, showing slightly better results on the optimal assignment magnitude error metrics.

\subsection{Additional validation on ATLAS molecular dynamics data}
\label{atlas}

To further assess our model's generalization capabilities, we conducted an additional validation experiment using molecular dynamics (MD) data from the ATLAS database. This experiment provided an independent test of PETIMOT's performance on high-quality data with distinct characteristics from our training data.

The ATLAS MD dataset underwent a systematic preprocessing pipeline. First, we extracted the principal components from the MD trajectories. Subsequently, we assigned samples to appropriate cross-validation folds, ensuring strict exclusion of both structural and sequential similarity between training and test sets. This rigorous assignment process yielded 400 samples suitable for evaluation. The inference was then performed by applying our trained PETIMOT models to predict the conformational motions for each sample.

\begin{figure}[ht]
     \centering
\includegraphics[width=0.75\linewidth]{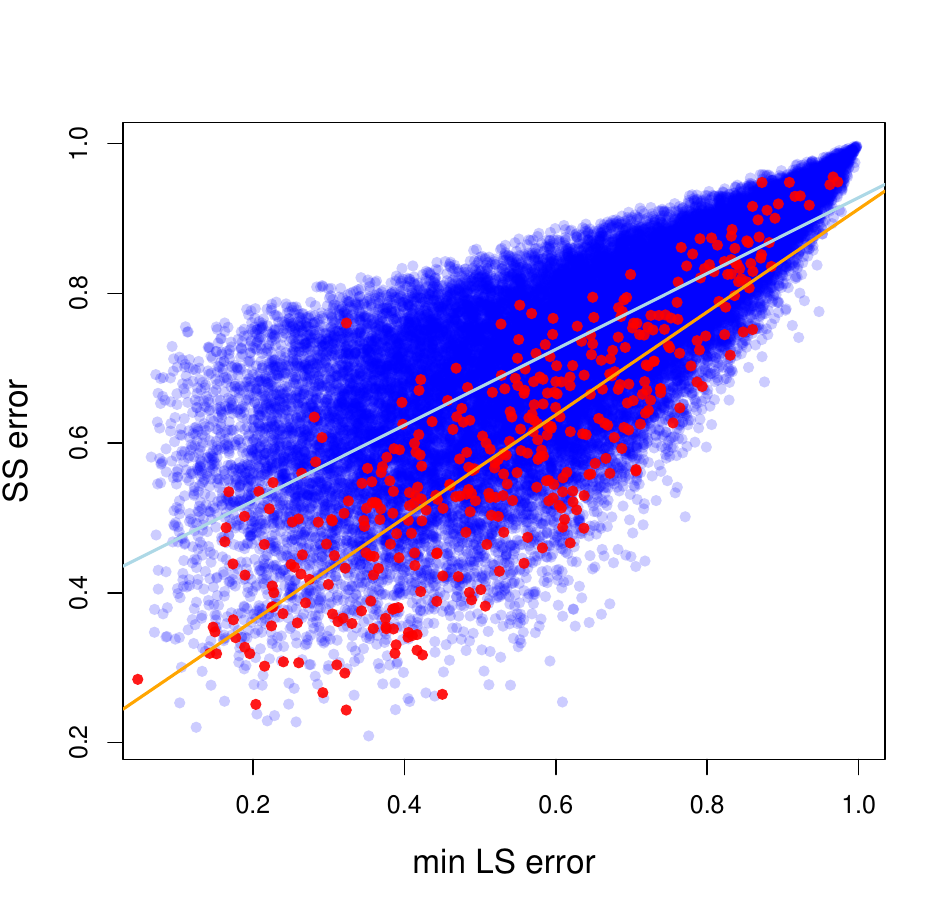}
\caption{{\bf SS error in function of Min. LS error.} Blue dots are samples from our PDB dataset while red dots correspond to the ATLAS MD samples.}
     \label{fig:minLS_SS}
 \end{figure}

\clearpage
\section{Additional results \label{resultsA}}

\subsection{Generalisation and robustness}

The performances achieved by PETIMOT-\emph{default} on \emph{test824} (Table \ref{tab:additional}, success rate of 43.57\%) generalise to the full dataset with PETIMOT-\emph{5folds} (Table \ref{tab:fullComp}, success rate of 38.98\%).  Moreover, PETIMOT consistently outperforms the NMA on both \emph{test824} and the full dataset.

\begin{table}[htb!]
\caption{{\bf Comparison of PETIMOT\emph{-5folds} with the NMA on the full dataset ($\sim$37k samples).} {For each sample, we evaluate the 4 motions predicted by PETIMOT against the 4 ground-truth motions. We allow the NMA more flexibility by considering either the 10 or the 4 lowest frequency modes.} Min. stands for the best matching pair of predicted and ground-truth vectors. OLA refers to the optimal linear assignment between all predicted and ground-truth vectors. Arrows indicate whether higher ($\uparrow$) or lower ($\downarrow$) metrics values are better. Best results are shown in \textbf{bold}.}
\centering
\begin{tabular}{lccc}
\toprule
Metrics & PETIMOT & \multicolumn{2}{c}{NMA}  \\
\midrule
        &         & first 10 & first 4 \\
        &         & modes & modes \\
\midrule
Success Rate (\%) $\uparrow$ & \textbf{38.98} & 25.43 & 24.40  \\
\midrule
Min. LS Error $\downarrow$ & \textbf{0.64 $\pm$ 0.20} & 0.71 $\pm$ 0.19 & 0.72 $\pm$ 0.20  \\
Min. Magnitude Error $\downarrow$ & \textbf{0.23 $\pm$ 0.12} & 0.24 $\pm$ 0.11 & 0.27 $\pm$ 0.14  \\
\midrule
OLA LS Error $\downarrow$ & \textbf{0.84 $\pm$ 0.09} & 0.85 $\pm$ 0.10 & 0.88 $\pm$ 0.09 \\
OLA Magnitude Error $\downarrow$ & 0.41 $\pm$ 0.13 & \textbf{0.38 $\pm$ 0.12} & 0.47 $\pm$ 0.15  \\
\midrule
Global SS Error $\downarrow$ & 0.75 $\pm$ 0.13 & \textbf{0.67 $\pm$ 0.17} & 0.79 $\pm$ 0.14  \\
\bottomrule
\end{tabular}
\label{tab:fullComp}
\end{table}

\begin{figure}[htb!]
    \centering
    \includegraphics[width=1\linewidth]{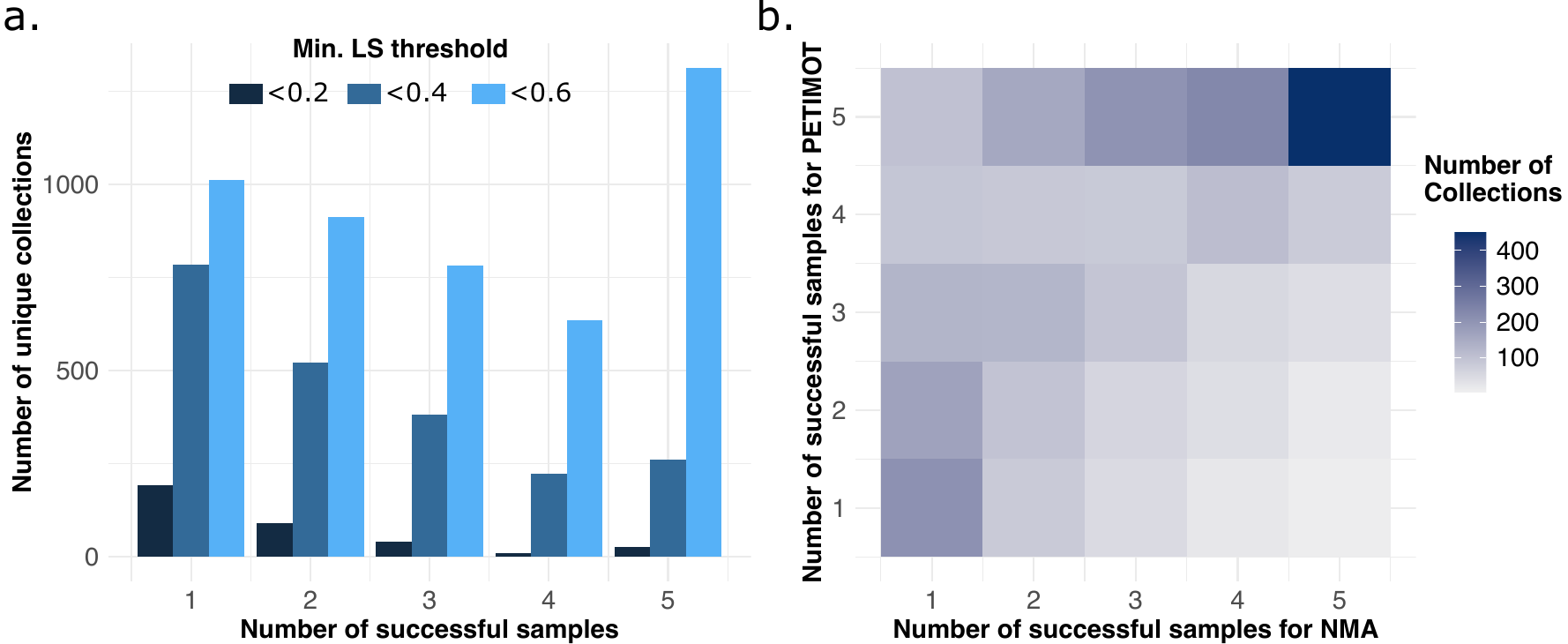}
    \caption{{{\bf Influence of the reference conformation on PETIMOT performance.} {\bf a.} Number of samples per conformational collection for which PETIMOT-\emph{5folds} predicted at least one acceptable motion, with LS below 0.2 (dark), 0.4 (mild), or 0.6 (light). The numbers of successful collections are 361 (4.92\% of 7,335 in total), 2,174 (29.64\%), and 4,658 (63.50\%), respectively. {\bf b.} Heatmap comparing the number of successful samples, with Min. LS below 0.6, for PETIMOT-\emph{5folds} (y-axis) and the NMA (x-axis). The number of successful collections for both methods is 2,822 (38.47\% of 7,335 in total).}}
    \label{supfig:influRef}
\end{figure}

{We further assessed the robustness of PETIMOT's performance across different reference conformations derived from the same experimental collection. When PETIMOT-\emph{5folds} predicted at least one acceptable motion, with Min. LS below 0.6, for a given collection, it did so for at least 2 out of 5 reference conformations in almost 80\% of cases (Fig. \ref{supfig:influRef}a). It predicted acceptable motions for all 5 reference conformations in 28\% of cases. These results suggest limited sensitivity of PETIMOT performance to the choice of reference conformation. Furthermore, PETIMOT-\emph{5folds} is more robust to the choice of reference conformation than the NMA (Fig. \ref{supfig:influRef}b). When both PETIMOT-\emph{5folds} and the NMA predicted at least one acceptable motion for a given collection, PETIMOT-\emph{5folds} did so for more conformations than the NMA in about half of the cases. The NMA predicted acceptable motions for more conformations in only 17\% of the cases.} 

\subsection{Comparison with other methods}

{Table~\ref{tab:additional} provide a comprehensive comparison of PETIMOT with AlphaFlow, BioEmu, and NMA variants on motion-related metrics. PETIMOT achieves the highest success rate (43.57\%) and consistently outperforms all baselines on Min. and OLA error metrics. The NMA with 10 modes remains competitive on OLA Magnitude Error and Global SS Error — metrics that aggregate over all predicted motions and may therefore favor methods with access to a larger mode set — while PETIMOT leads on all metrics directly reflecting the quality of individual motion predictions. Across all methods, PETIMOT is also by far the fastest, running in under 16 seconds per protein compared to over 38 hours for AlphaFlow and BioEmu.}

\begin{table}[bht!]
\caption{{\bf PETIMOT's performance on the test set and comparison with other methods.}  PETIMOT\emph{-default} is compared with AlphaFlow, BioEmu, and the NMA on \emph{test824}. {For each protein, we evaluate the 4 motions predicted by PETIMOT or inferred from AlphaFlow/BioEmu predicted ensembles against the 4 ground-truth motions. We allow the NMA more flexibility by considering the 10 lowest frequency modes.} Min. stands for the best matching pair of predicted and ground-truth motions. OLA refers to the optimal linear assignment between all predicted and ground-truth vectors. Arrows indicate whether higher ($\uparrow$) or lower ($\downarrow$) values are better, best results highlighted in \textbf{bold}.
{The running times were measured on an Intel(R) Xeon(R) W-2245 CPU @ 3.90GHz equipped with GeForce RTX 3090 for PETIMOT, Alpha/ESM-Flow, and the NMA, and on an AMD Ryzen 9 7950X 16-Core CPU @ 5.88GHz equipped with NVIDIA RTX A6000 for BioEmu, which can be up to 30\% faster in some tasks.}
}
\centering
\begin{tabular}{lccccc}
\toprule
Metrics & PETIMOT & AlphaFlow & BioEmu & \multicolumn{2}{c}{NMA} \\
\midrule
        &  &  &  & first 10 & first 4 \\
        &  &  &   & modes & modes \\
\toprule
Running time $\downarrow$ & \textbf{15.82s} & 38h 07min  & 39h12min & \multicolumn{2}{c}{43.59s} \\
\toprule
Success Rate (\%) $\uparrow$ &  \textbf{43.57} & 31.80 & 31.34 & 25.73 & 24.88 \\
\midrule
Min. LS Error $\downarrow$ & \textbf{0.61 $\pm$ 0.22} & 0.68 $\pm$ 0.21 & 0.68 $\pm$ 0.20 & 0.70 $\pm$ 0.19 & 0.72 $\pm$ 0.20 \\
Min. Magnitude Error $\downarrow$ & \textbf{0.21 $\pm$ 0.12} & 0.24 $\pm$ 0.12 & 0.23 $\pm$ 0.12 & 0.25 $\pm$ 0.11 & 0.27 $\pm$ 0.14 \\
\midrule
OLA LS Error $\downarrow$ & \textbf{0.83 $\pm$ 0.10} & 0.86 $\pm$ 0.10 & 0.86 $\pm$ 0.10 & 0.85 $\pm$ 0.10 & 0.88 $\pm$ 0.10 \\
OLA Magnitude Error $\downarrow$ & 0.41 $\pm$ 0.14 & 0.43 $\pm$ 0.14 & 0.42 $\pm$ 0.13 & \textbf{0.39 $\pm$ 0.12} & 0.48 $\pm$ 0.15 \\
\midrule
Global SS Error $\downarrow$ & 0.73 $\pm$ 0.14 & 0.78 $\pm$ 0.14 & 0.77 $\pm$ 0.14 & \textbf{0.67 $\pm$ 0.16} & 0.79 $\pm$ 0.14 \\
\bottomrule
\end{tabular}
\label{tab:additional}
\end{table}

\subsection{Comparison with NMA variants}

{We compared PETIMOT with several NMA variants (Table \ref{tab:NMAsweeps}). We explored several distance cutoffs to define the Elastic Network Model (ENM). The cutoff of 10\AA{} yields the best performance (Success Rate 25.73\%, Min. LS Error 0.70), while smaller (7.5\AA{}) and larger (13\AA{}) cutoffs perform slightly worse, suggesting a moderate sensitivity to this hyperparameter. In addition, we increased the resolution of the ENM by including all atoms. For evaluation, all-atom NMA displacement vectors are assessed at C$\alpha$ positions only, ensuring that metrics remain directly comparable across all methods. This variant achieves the best NMA performance overall on several metrics, including Success Rate (28.52\%) and OLA Magnitude Error (0.37), as well as Global SS Error (0.66). However, the differences with respect to the best performing C$\alpha$-based NMA (at 10\AA{} cutoff) remain small and hence, increasing the ENM resolution does not affect the conclusions of our evaluation. PETIMOT outperforms all NMA variants on the three primary metrics — Success Rate (43.57\%), Min. LS Error (0.61) and Min. Magnitude Error (0.21) — with consistent margins, while the NMA variants perform competitively on OLA Magnitude Error and Global SS metrics. This is consistent with the asymmetry in the comparison setup: NMA is allowed 10 modes while PETIMOT predicts only 4, which may favor NMA on metrics that aggregate over all predicted motions.}

\begin{table}[bht!]
\caption{{{\bf Comparison between PETIMOT and NMA variants.}  PETIMOT\emph{-default} is compared with NMA variants on \emph{test824}. For each sample, we evaluate the 4 motions predicted by PETIMOT against the 4 ground-truth motions. We allow the NMA more flexibility by considering the 10 lowest frequency modes. We vary the distance cutoff used to defined the elastic network model and its resolution (C$\alpha$ atoms only or all atoms) across NMA variants. Min. stands for the best matching pair of predicted and ground-truth vectors. OLA refers to the optimal linear assignment between all predicted and ground-truth vectors. Arrows indicate whether higher ($\uparrow$) or lower ($\downarrow$) metrics values are better. Best results are shown in \textbf{bold}.}}
\centering
\begin{tabular}{|l|c|c|c|c|c|}
\hline
Method & PETIMOT & \multicolumn{4}{c|}{NMA} \\
\hline
Resolution        &     {C$\alpha$ only}    &  \multicolumn{3}{c|}{C$\alpha$ only} & All-atom \\
\hline
Cutoff distance &         & 7.5 \AA{} & 10 \AA{} & 13 \AA{} & 5 \AA{}\\
\hline
Success Rate (\%) $\uparrow$ &  \textbf{43.57} & 19.05 & 25.73 & 24.27 & 28.52 \\
\hline
Min. LS Error $\downarrow$ & \textbf{0.61 $\pm$ 0.22} & 0.75 $\pm$ 0.19 & 0.70 $\pm$ 0.19 & 0.71 $\pm$ 0.19 & 0.69 $\pm$ 0.19\\
Min. Magnitude Error $\downarrow$ & \textbf{0.21 $\pm$ 0.12} & 0.28 $\pm$ 0.12 & 0.25 $\pm$ 0.11 & 0.25 $\pm$ 0.12 & 0.23 $\pm$ 0.11\\
\hline
OLA LS Error $\downarrow$ & \textbf{0.83 $\pm$ 0.10} & 0.88 $\pm$ 0.09 & 0.85 $\pm$ 0.10 & 0.85 $\pm$ 0.10 & 0.84 $\pm$ 0.10\\
OLA Magnitude Error $\downarrow$ & 0.41 $\pm$ 0.14 & 0.42 $\pm$ 0.13 & 0.39 $\pm$ 0.12 & 0.40 $\pm$ 0.13 & \textbf{0.37 $\pm$ 0.12}\\
\hline
Global SS Error $\downarrow$ & 0.73 $\pm$ 0.14 & 0.73 $\pm$ 0.16 & 0.67 $\pm$ 0.16 & 0.68 $\pm$ 0.17 & \textbf{0.66 $\pm$ 0.16}\\
\hline
\end{tabular}
\label{tab:NMAsweeps}
\end{table}

\subsection{Influence of the number of predicted components.}

\begin{figure}[htb!]
    \centering
    \includegraphics[width=0.8\linewidth]{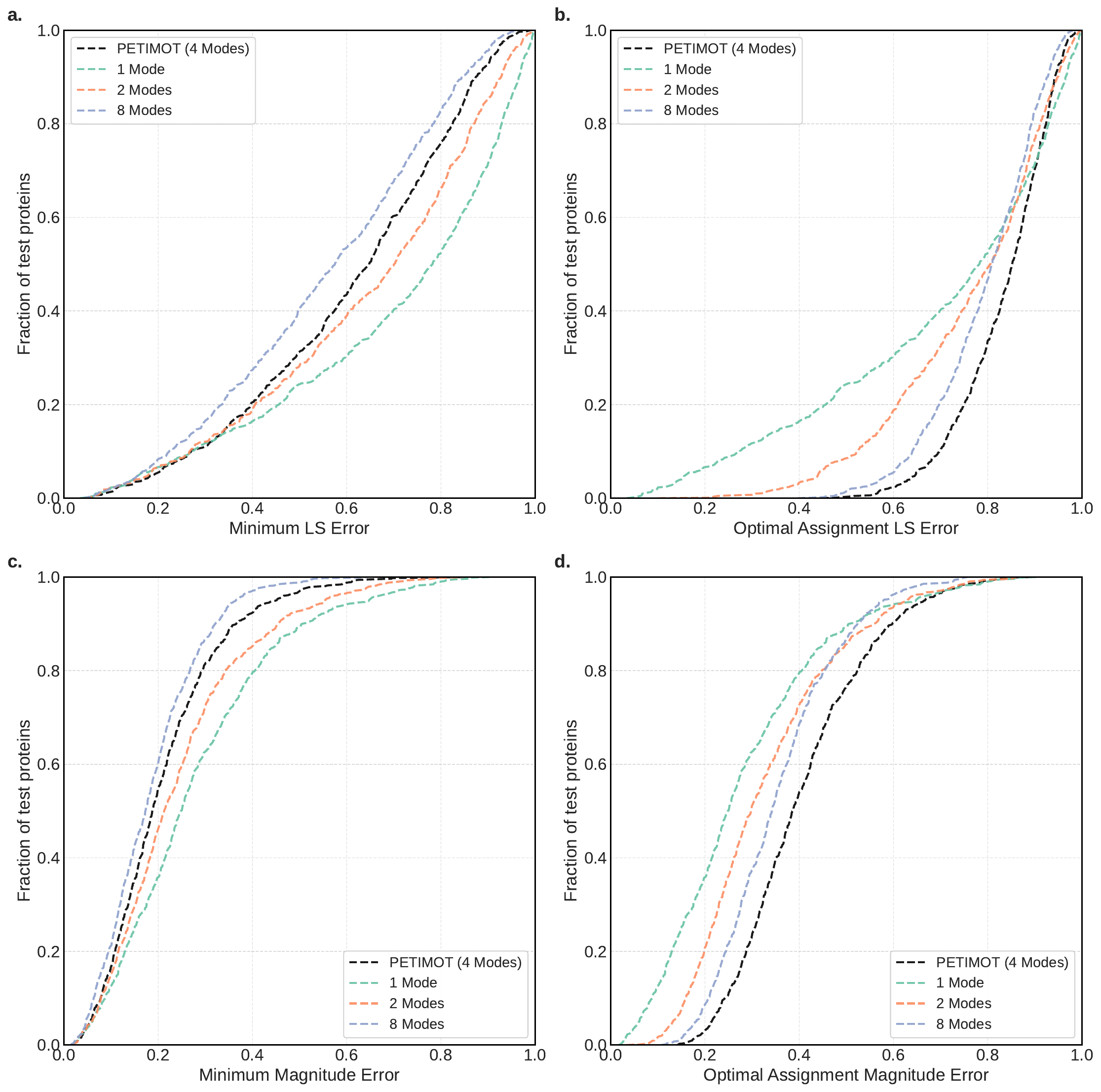}
    \caption{{\bf Impact of the number of predicted components.} We report cumulative curves for LS error (a-b) and magnitude error (c-d). For each protein, we computed the error either for the best-matching pair of predicted and ground-truth vectors (a,c) or for the best combination of all pairs of predicted and ground-truth vectors using optimal linear assignment (b,d). We compare models trained to predict different numbers of components (modes): 1, 2, 4, or 8, using only the LS loss.}
    \label{fig:ablation:modes}
\end{figure}

We also experimented with a different number of predicted components, while maintaining the number of ground-truth components fixed at $L=4$. 
For these experiments, we trained additional models with the LS loss only:
\begin{itemize}
\item Single component prediction (1 mode).
\item Reduced component prediction (2 modes).
\item Extended component prediction (8 modes).
\end{itemize}
We compare these with our default setting of $K=4$ predicted components.
Figure \ref{fig:ablation:modes} shows the results. The key insight is that minimum-based metrics (Fig. \ref{fig:ablation:modes}a,c) and assignment-based metrics (Fig. \ref{fig:ablation:modes}b,d) measure different aspects of subspace quality:
Minimum metrics measure the best possible match between any predicted and ground-truth component. These improve with more predicted components (from 1 to 8) because having more candidates increases the likelihood of finding at least one good match with each ground-truth component.
Optimal assignment metrics measure overall subspace alignment by finding the best one-to-one matching between predicted and ground-truth components. Here, models with fewer predicted components (1-2) perform better because they face fewer constraints in the assignment problem - each predicted component can be matched to the best available ground-truth component without competition.
The 8-component model maintains the best performance overall, as having more candidate vectors provides flexibility while still capturing the 4-dimensional ground-truth subspace effectively.

\begin{figure}[thb!]
    \centering
    \includegraphics[width=0.8\linewidth]{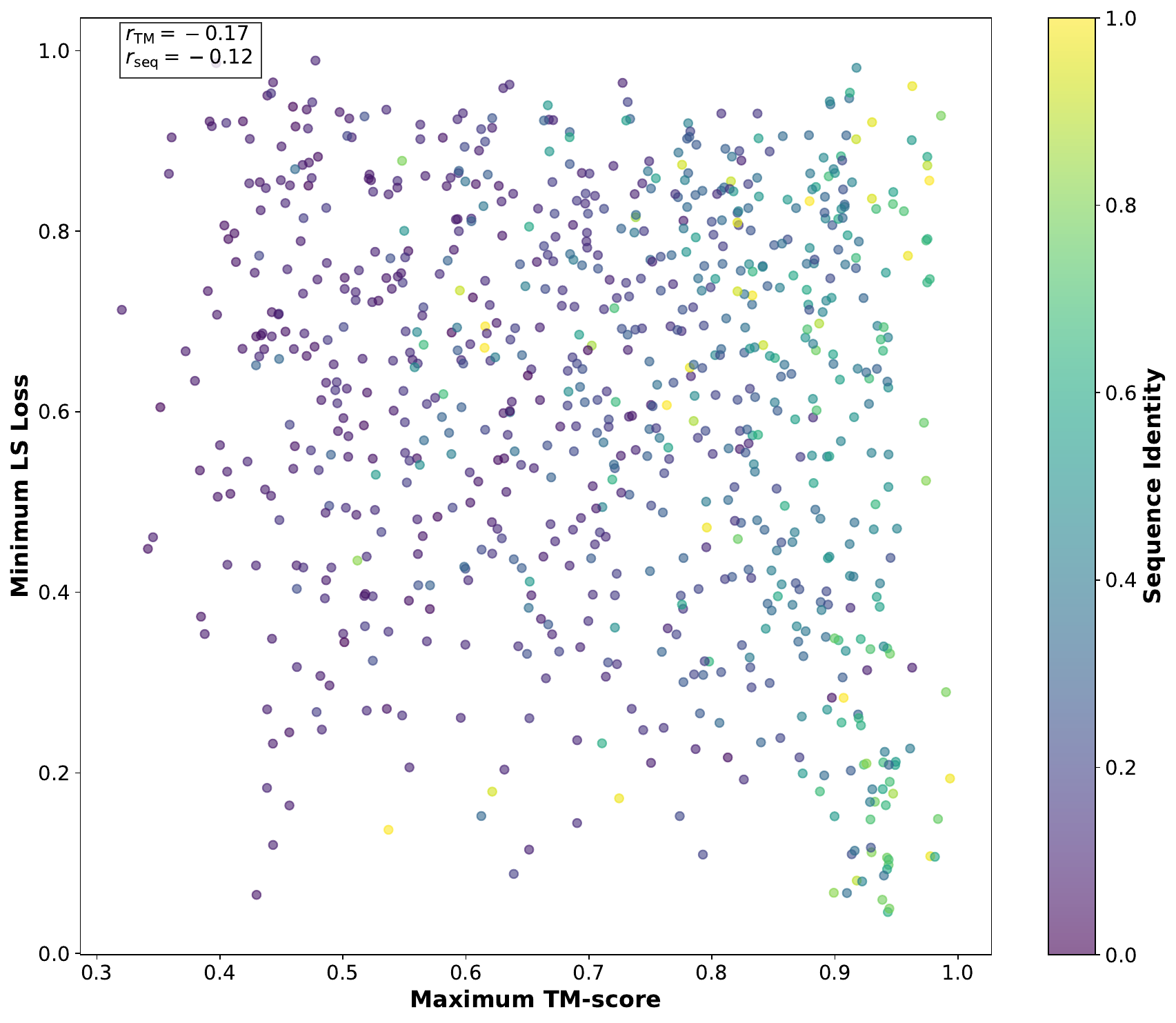}
    \caption{{\bf Relationship between PETIMOT's prediction accuracy and structural/sequence similarity with the training set.} The minimum LS error is plotted against the maximum TM-score between each test protein and any protein in the training set. Points are colored by the maximum sequence identity to the training samples.}
    \label{fig:tm_seq_sim}
\end{figure}

Figure \ref{fig:tm_seq_sim} compares the accuracy of the predicted test proteins (minimum LS loss) with the structural (TM-score) and sequence (sequence identity) distances to the training set. We do not see a clear correlation between the prediction accuracy and the similarity to the training examples. Please also see Fig. \ref{fig:indiv}b-c for comparison.

\subsection{Visualisation of the predictions}

\begin{figure}[t]
    \centering
    \includegraphics[width=0.8\linewidth]{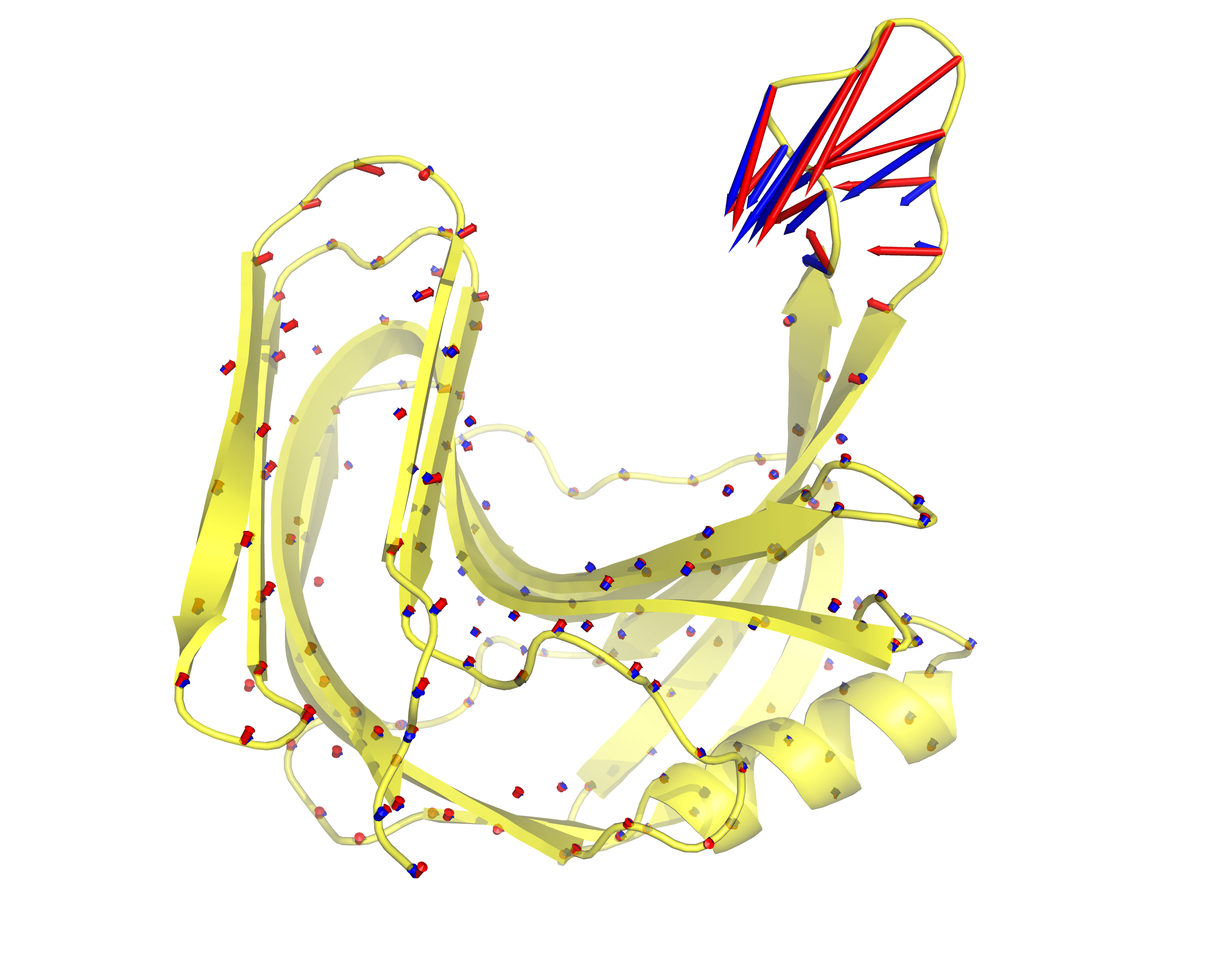}
    \caption{{\bf Visualization of predicted (blue arrows) and ground-truth (red arrows) motion vectors for PDB structure 3EXU (chain A), with LS error of 0.20.} The predicted deformation was used to generate the interpolated conformations shown in Fig. \ref{fig:indiv}b.}
    \label{fig:3EXUA:arrow}
\end{figure}

\begin{figure}[t!]
    \centering
    \includegraphics[width=0.8\linewidth]{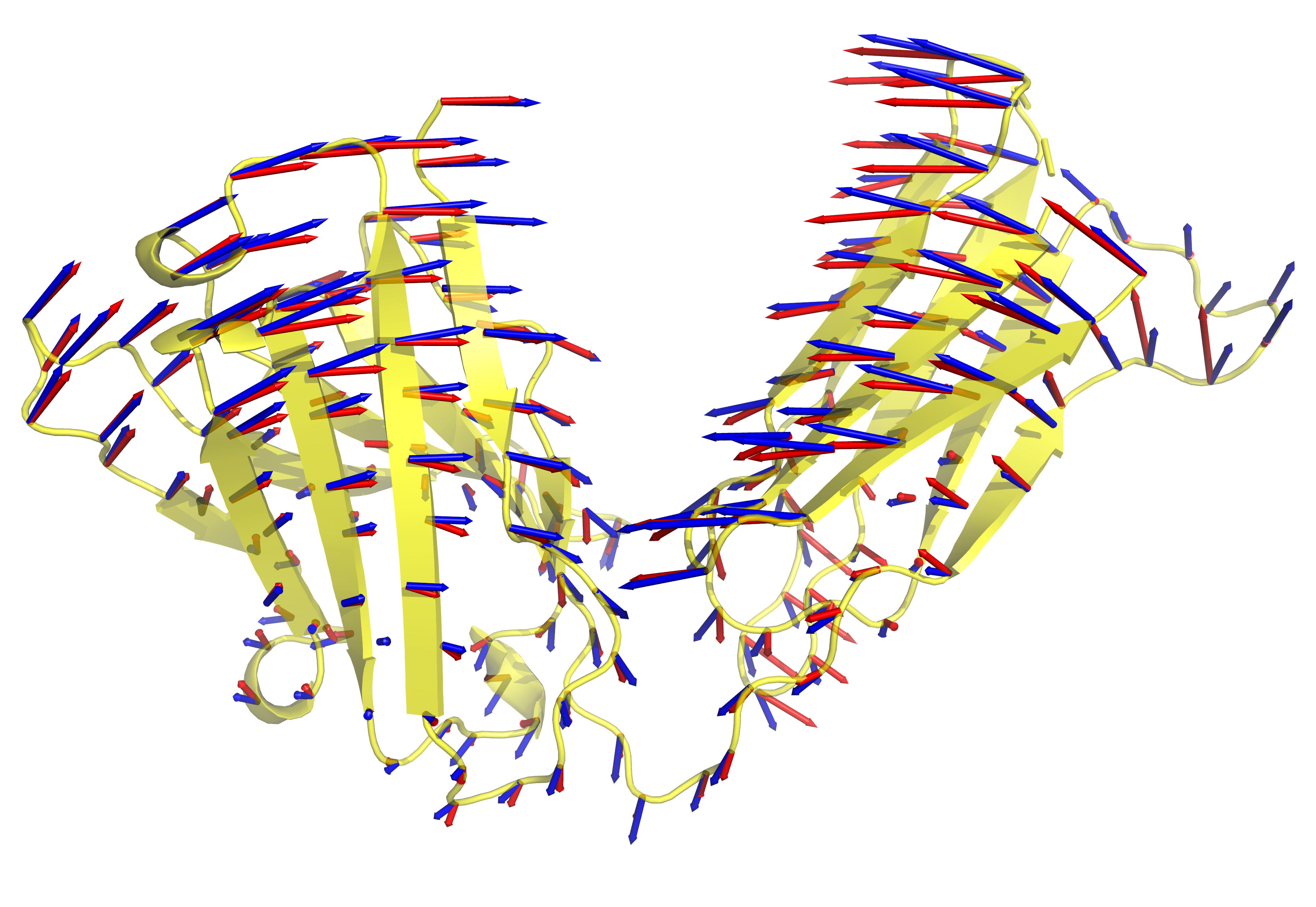}
    \caption{{\bf Visualization of predicted (blue arrows) and ground-truth (red arrows) motion vectors for PDB structure 7SD2, with LS error of 0.18.} The predicted deformation was used to generate the interpolated conformations shown in Fig. \ref{fig:indiv}c.}
    \label{fig:7SD2A:arrow}
\end{figure}

Figures \ref{fig:3EXUA:arrow} and \ref{fig:7SD2A:arrow} show predicted (blue arrows) and ground-truth (red arrows) motion vectors for the xylanase A from {\it Bacillus subtilis} and the periplasmic domain of Gliding motility protein GldM from {\it Capnocytophaga canimorsus}, respectively.

\begin{figure}[t!]
    \centering
    \includegraphics[width=0.8\linewidth]{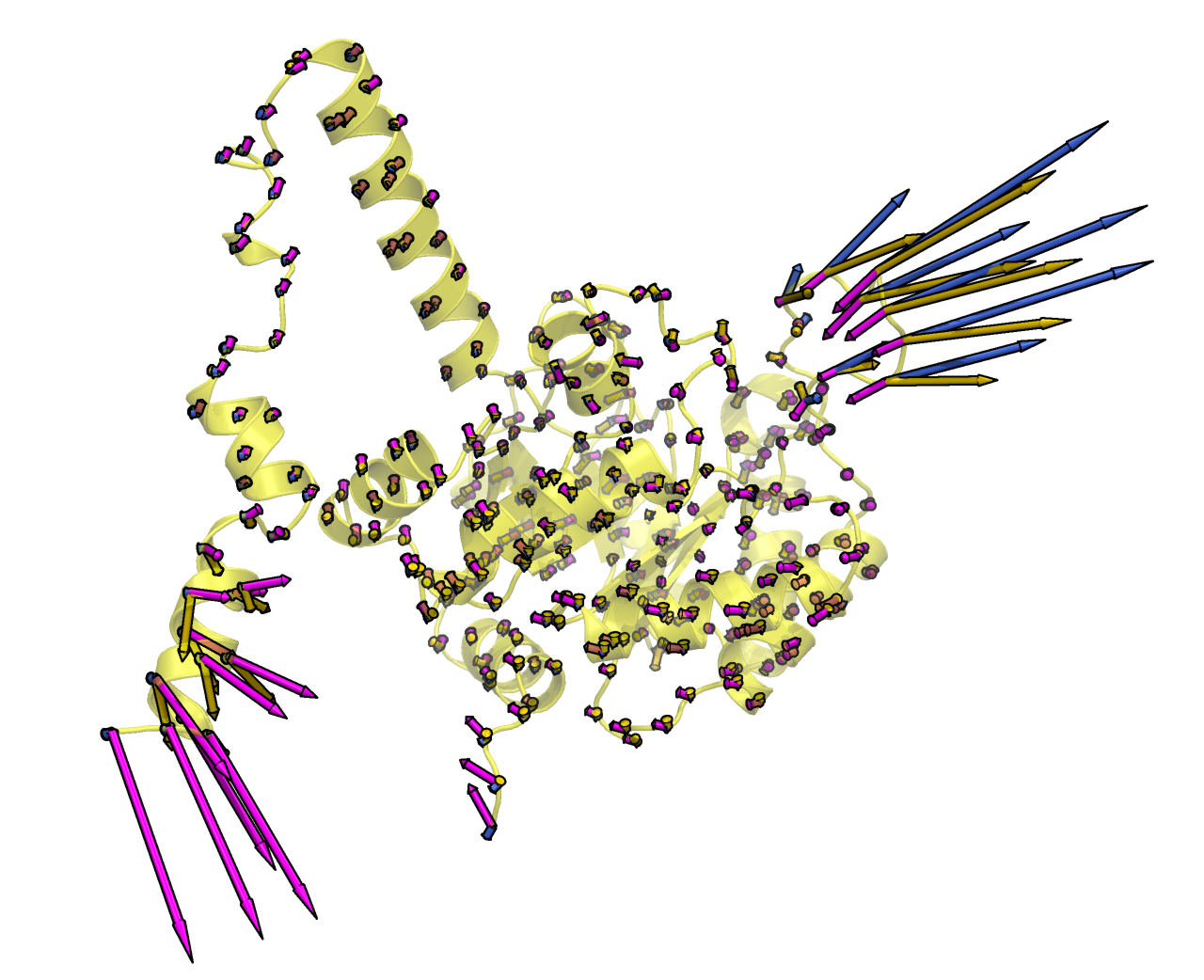}
    \caption{
    {\bf Visualization of ground-truth and predicted motion vectors for PDB structure 1OQF (chain A).} Ground-truth main motions is depicted as yellow arrows. The best-matching motions predicted by PETIMOT (fourth) and BioEmu (first) are coloured in blue and magenta, respectively.
    }
    \label{fig:1OQFA:arrow}
\end{figure}

Figure \ref{fig:1OQFA:arrow} show the ground-truth main motion (yellow arrows) exhibited by experimental structures of the enzyme 2-methylisocitrate lyase (PrpB) from {\it Escherichia coli} and the best-matching predictions from PETIMOT (blue) and BioEmu (magenta). This enzyme, which catalyses the last step of the methylcitrate cycle, belongs to the isocitrate lyase protein family. Members of this family share an opening-closing loop mechanism associated with ligand binding. PETIMOT captures this loop motion very precisely (Min. LS error below 0.4) while the BioEmu ensemble mostly exhibits motions of the C-terminal helix.

\begin{figure}[t]
    \centering
    \includegraphics[width=0.8\linewidth]{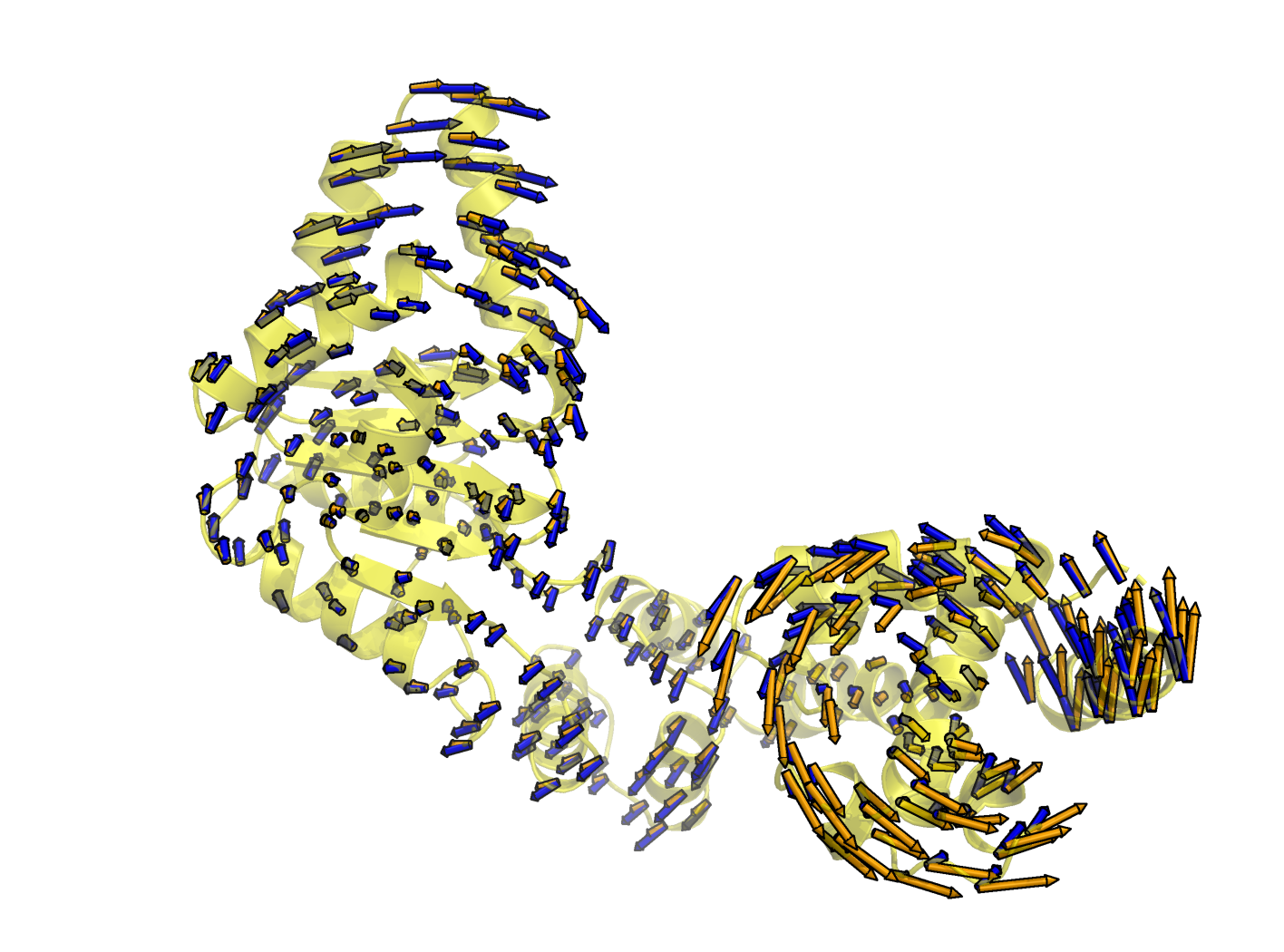}
    \caption{
    {\bf Visualization of predicted motion vectors.} PETIMOT is in blue (Min. LS = 0.73) and the NMA is in orange (Min. LS = 0.30) for the PDB structure 2HCB (chain C).
    }
    \label{fig:2HCBC:arrow}
\end{figure}

\clearpage
\section{Licenses for used resources}
\label{licences}

In this work, we use several existing resources. The protein structures were obtained from the Protein Data Bank (PDB, https://www.rcsb.org/, version accessed on June 2023) which is distributed under the CC0 1.0 Universal Public Domain Dedication license (CC0 1.0). We complemented PDB data with data from PDB-redo (accessed June 2023) developed by Joosten {\it et al.} \cite{joosten2014pdb_redo}, available at \url{https://pdb-redo.eu} under the licence specified at \url{https://pdb-redo.eu/license}. For protein language modeling, we employed ProstT5 developed by Heinzinger {\it et al.} \cite{heinzinger2023prostt5}, available under the MIT license at \url{https://huggingface.co/Rostlab/ProstT5}, and ESM-Cambrian 600M (verison esmc-600m-2024-12) developed by the EvolutionaryScale Team \cite{esm2024cambrian}, available under the Cambrian Non-Commercial License at \url{https://huggingface.co/EvolutionaryScale/esmc-600m-2024-12}. Additional resources include the DANCE method (version of Oct 8, 2024) developed by Lombard {\it et al.}, available under the MIT license at \url{https://github.com/PhyloSofS-Team/DANCE}. As baselines, we ran the NOLB method (version 1.9) developed by Hoffmann {\it et al.} \cite{hoffmann2017nolb} and available at \url{https://team.inria.fr/nano-d/software/nolb-normal-modes/}, the AlphaFlow (version AlphaFlow-PDB distilled) and ESMFlow (version ESMFlow-PDB distilled) models developed by Jing {\it et al.} \cite{jing2024alphafold} and available at \url{https://github.com/bjing2016/alphaflow},
and BioEmu  developed by Lewis {\it et al.} \cite{lewis2025scalable} available under the MIT license at \url{https://github.com/microsoft/bioemu}.
We used TM-align (version 20220412) developed by Zhang and Skolnick \cite{zhang2005tm} and available at \url{https://zhanggroup.org/TM-align/} to perform all-to-all pairwise structural alignments between train and test protein conformations and compute TM-scores.

\begin{funding}
This work has been funded by the European Union (ERC, PROMISE, 101087830). Views and opinions expressed are however those of the author(s) only and do not necessarily reflect those of the European Union or the European Research Council. Neither the European Union nor the granting authority can be held responsible for them. For the purpose of Open Access, a CC-BY public copyright licence has been applied by the authors to the present document and will be applied to all subsequent versions up to the Author Accepted Manuscript arising from this submission.
\end{funding}

\subsubsection*{Author Contributions}
Conceptualisation: S.G., E.L. Data curation: V.L. Formal analysis: S.G, with feedback from E.L. and V.L. Funding acquisition: E.L. Investigation: all authors. Methodology, V.L. with the help of J.N.V. Software: V.L. Supervision: S.G. and E.L. Validation: all authors. Visualisation: all authors. Writing – original draft: V.L., S.G., and E.L. Writing – review \& editing: all authors. 

\ConflictsOfInterest{All authors declare that they have no conflicts of interest.
}

\DataAvailability{
The code and the data are available at \url{ https://github.com/PhyloSofS-Team/PETIMOT}.
}

\clearpage

\bibliography{petitmot}

\end{document}